\documentclass[
aps,
reprint,
superscriptaddress,
longbibliography,
nofootinbib
]{revtex4-2}

\usepackage[ruled,vlined,linesnumbered]{algorithm2e}

\usepackage[T1]{fontenc}
\usepackage[utf8]{inputenc}
\usepackage{amsmath,amssymb,amsfonts,amsthm,bm}
\usepackage{physics}
\usepackage{bbm}
\usepackage{slashed}
\usepackage{graphicx}
\usepackage[dvipsnames]{xcolor}
\usepackage{tikz}
\usepackage{tabularx}
\usepackage{multirow}
\usepackage{array}
\usepackage{diagbox}
\usepackage{multibib}

\newcites{supp}{References}
\usepackage[colorlinks=true,citecolor=blue,linkcolor=blue,urlcolor=blue]{hyperref}

\definecolor{mycolor}{rgb}{0.8, 0.4, 0.4}

\DeclareMathOperator*{\argmin}{arg\,min}

\begin{document}
	
	\title{
		Variational preparation of thermofield double states for SYK models via multi-angle QAOA: sequential angle pruning for circuit reduction
	}
	
	\author{Haji Muhammad Husnain Ashfaq}
	\email{m.husnainashfaq2911@gmail.com}
	\affiliation{Department of Physics and Photon Science, Gwangju Institute of Science and Technology,\\123 Cheomdan-gwagiro, Gwangju 61005, Republic of Korea}
	
	\author{Moongul Byun}
	\email{moongulbyun@gm.gist.ac.kr}
	\affiliation{Department of Physics and Photon Science, Gwangju Institute of Science and Technology,\\123 Cheomdan-gwagiro, Gwangju 61005, Republic of Korea}
	
	\author{Keun-Young Kim}
	\email{fortoe@gist.ac.kr}
	\affiliation{Department of Physics and Photon Science, Gwangju Institute of Science and Technology,\\123 Cheomdan-gwagiro, Gwangju 61005, Republic of Korea}
	
	\begin{abstract}
		Variational preparation of thermofield double (TFD) states can require deep quantum circuits, particularly for interacting many-body systems.
		Reducing these circuits while retaining high fidelity is therefore crucial for TFD-state preparation on noisy quantum processors.
		We study this problem by applying the multi-angle quantum approximate optimization algorithm (ma-QAOA) to TFD-state preparation and introducing two top-down sequential angle-pruning algorithms.
		Starting from the optimized initial ma-QAOA circuit, both algorithms sequentially remove Pauli-string evolutions with small optimized angles and reoptimize the remaining parameters after each removal.
		We apply these algorithms to Gaussian and binary Sachdev--Ye--Kitaev (SYK) models in both dense and sparse cases.
		We find that ma-QAOA prepares the target TFD states with high fidelity and that sequential small-angle pruning retains high fidelity while reducing the circuit depth, particularly at low temperature.
		Moreover, using the post-reoptimization cost in sequential small-angle pruning further improves the fidelity.
		For the binary sparse $N=10$ SYK model at $\beta=10$, $88.8\%$--$92.1\%$ of the nonlocal Pauli-string evolutions are removed while retaining an average fidelity of approximately $95\%$.
		Finally, we propose extensions of the sequential pruning algorithms toward quantum--classical hybrid implementation.
	\end{abstract}
	
	\maketitle
	
	\section{Introduction and summary}
	\label{sec:intro}
	
	The thermofield double (TFD) state is a canonical purification of a Gibbs state.
	It is a finite-temperature entangled state of two copies of a quantum many-body system and has been used in quantum simulations of traversable-wormhole protocols~\cite{Jafferis2022,Byun:2026ewk} in the context of holographic duality~\cite{Maldacena:2018lmt}.
	Thus, accurate preparation of the TFD state on a quantum processor is important because its fidelity directly affects the experimental results.
	
	Exact preparation of the TFD state on a quantum processor is challenging because its direct construction involves nonunitary imaginary-time evolution.
	One way to prepare the TFD state is to employ variational quantum algorithms (VQAs), which approximate the TFD state using a variational quantum circuit optimized with a suitable objective function~\cite{McClean_2016,Cerezo2021}.
	One standard choice of this function is the energy expectation value, as in the variational quantum eigensolver (VQE), which approximates the ground state of a Hamiltonian~\cite{Peruzzo2014,Kandala2017}.
	For TFD-state preparation via VQE, one can use an engineered Hamiltonian whose ground state approximates the TFD state~\cite{PhysRevA.104.012427,PhysRevA.111.012432}.
	Other objective functions have also been considered for TFD-state preparation, including engineered cost functions~\cite{Premaratne:9259931,Sagastizabal2021} and infidelity in numerical simulations~\cite{PhysRevLett.123.220502}.
	
	Beyond the choice of objective function, the performance of a VQA also depends on the structure of the variational quantum circuit~\cite{PhysRevResearch.2.023074,Ostaszewski2021structure,Du2022,PhysRevResearch.6.033033}.
	A Hamiltonian-informed ansatz provides a simple way to construct the variational quantum circuit using Hamiltonians related to the target system.
	A representative example is the quantum approximate optimization algorithm (QAOA)~\cite{Farhi2014qaoa,Hadfield2019}.
	In its standard form, QAOA alternates unitary evolutions generated by two noncommuting Hamiltonians, referred to as the cost and mixer Hamiltonians.
	As a hybrid quantum--classical algorithm, QAOA can provide a time-efficient approach to approximate state preparation with shallow circuits, making it attractive for near-term quantum simulation of many-body systems~\cite{Farhi:2016zwi,Preskill2018quantumcomputingin,PhysRevA.99.052332,SciPostPhys.6.3.029,PhysRevX.10.021067,doi:10.1073/pnas.2006373117,PRXQuantum.2.010309}.
	QAOA-motivated approaches have also been applied to TFD-state preparation using intra- and inter-system Hamiltonians~\cite{PhysRevLett.123.220502,Premaratne:9259931,Zhu:2020,Sagastizabal2021}.
	
	However, for complex problems, QAOA can require many gates and large circuit depth, limiting its implementation on near-term quantum hardware~\cite{10.1007/s11128-021-03001-7,Harrigan2021,Vijendran_2024}.
	One approach that can enable circuit reduction is the multi-angle QAOA (ma-QAOA)~\cite{Herrman2022,Gaidai2024}.
	Unlike standard QAOA, ma-QAOA assigns independent variational parameters to individual Hamiltonian terms.
	In Ref.~\cite{Herrman2022}, many optimized parameters were found to be zero, so the corresponding gates could be removed while retaining the solution quality.
	
	Structurally similar to ma-QAOA, a variational circuit ansatz with independent parameters for different Hamiltonian terms has been applied to TFD-state preparation~\cite{Zhu:2020,Sagastizabal2021,PhysRevA.111.012432}.
	This suggests that ma-QAOA can also provide a natural ansatz for TFD-state preparation for many-body systems.
	At the same time, TFD-state preparation can require substantial circuit resources, making circuit reduction important for near-term quantum simulation~\cite{PhysRevLett.123.220502,PhysRevA.111.012432}.
	However, the application of ma-QAOA to TFD-state preparation together with systematic circuit reduction has not yet been explored; this motivates the present work.
	
	As a model for a many-body system, we consider the Sachdev--Ye--Kitaev (SYK) model, which describes strongly interacting $N$ Majorana fermions~\cite{PhysRevLett.70.3339,Kitaev2015}.
	In the large-$N$ and low-temperature limit, the SYK model shows maximal quantum chaos and is holographically dual to nearly $\mathrm{AdS}_2$ gravity~\cite{Maldacena2016,PhysRevD.94.106002,PhysRevLett.126.030602,10.1093/ptep/ptw124,PhysRevLett.117.111601}.
	These features make the model a useful platform for quantum simulations of traversable-wormhole protocols, which require accurate TFD-state preparation~\cite{Jafferis2022,Byun:2026ewk}.
	One can further consider sparse SYK models, which reduce the number of interaction terms while preserving quantum chaos~\cite{xu2020sparsemodelquantumholography,PhysRevD.103.106002,Caceres2021,PhysRevB.107.L081103,Orman2025}.
	This makes sparse SYK models suitable for near-term quantum simulations~\cite{Granet2026,Byun:2026ewk,Bang:2026eof}.
	
	There have been several variational approaches to thermal-state preparation for both dense and sparse SYK models~\cite{PhysRevA.100.032107,PhysRevA.104.012427,Araz:2024xkw,Kundu_2025}.
	Extending these works, we consider TFD-state preparation for the SYK models using ma-QAOA.
	The cost and mixer Hamiltonians are given by the two-sided SYK Hamiltonian and a left--right interaction, respectively.
	An independent variational angle is assigned to the Pauli-string evolution corresponding to each Hamiltonian term.
	
	However, since SYK Hamiltonians contain many interaction terms, the ma-QAOA ansatz correspondingly contains a large number of gates.
	Although zero-angle gates can be removed directly, a substantial number of gates can still remain, and implementing them can degrade the quantum simulation results.
	It therefore remains unclear \textit{how to systematically and substantially reduce these gates in ma-QAOA while retaining high fidelity for TFD-state preparation on quantum hardware}.
	
	To resolve this, we note that pruning redundant or weakly contributing circuit elements is a common strategy for compacting variational circuits~\cite{Sim_2021,Wang2022,Kulshrestha:2024qve,10.1021/acs.jctc.5c00535,Escofet2026,He:2026utf}.
	Previous ma-QAOA studies have also considered removing zero or near-zero variational angles for circuit reduction~\cite{Herrman2022,Kim:2026pnn}.
	These observations motivate a top-down pruning approach starting from the initial ma-QAOA ansatz.
	In addition, unlike bottom-up adaptive construction, this approach avoids the repeated search over an operator pool~\cite{Warren:2022evv}.
	We therefore apply this top-down pruning approach to TFD-state preparation with ma-QAOA; this is the main focus of this work.
	
\begin{figure*}[tb]
		\centering
		\includegraphics[width=\linewidth]{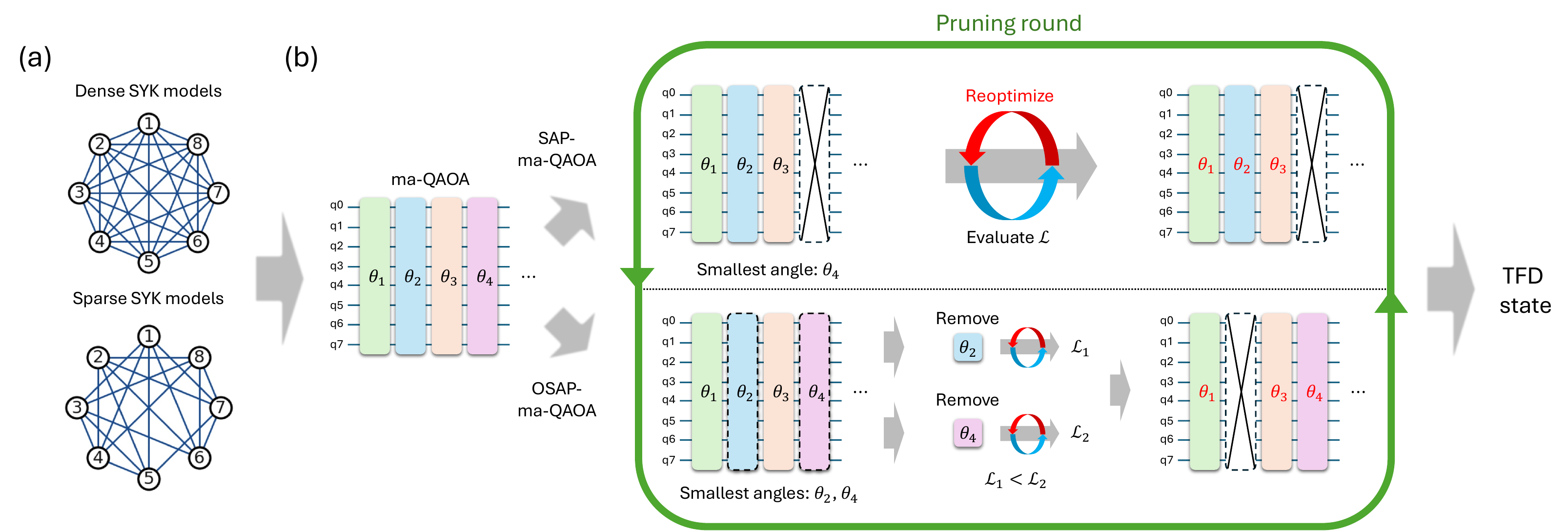}
		\caption{
			\label{figure1}
			Schematic overview of the SYK models and the circuit-pruning procedures in ma-QAOA.
			(a) Connectivity structures of the dense and sparse SYK models.
			(b) In SAP-ma-QAOA, the nonlocal Pauli-string evolution with the smallest angle magnitude (dashed) is removed and the remaining parameters are classically reoptimized.
			In OSAP-ma-QAOA, several evolutions with small angle magnitudes (dashed) are considered as removal candidates, and the candidate yielding the lowest post-reoptimization cost is selected; in the example shown, $\mathcal{L}_{1}<\mathcal{L}_{2}$.
			The colored gates represent different Pauli-string evolutions.
			The red and blue arrows indicate classical and quantum computations, respectively, and the angles shown in red are those updated by the reoptimization.
			The green loop indicates that the pruning-and-reoptimization procedure is repeated until the target number of blocks is reached.
		}
	\end{figure*}

	Specifically, we propose \textit{sequential angle-pruning ma-QAOA} (SAP-ma-QAOA).
	After optimizing the initial ma-QAOA circuit, this algorithm removes the nonlocal Pauli-string evolution with the smallest angle magnitude and classically reoptimizes the remaining parameters after each removal.\footnote{A related reoptimization procedure is used in recursive QAOA, where QAOA is repeatedly optimized after reducing the problem size~\cite{PhysRevLett.125.260505}.}
	We further propose \textit{objective-aware SAP-ma-QAOA} (OSAP-ma-QAOA).
	This algorithm considers several candidate Pauli-string evolutions with small optimized angles.
	Then, the evolution whose removal yields the lowest post-reoptimization cost is selected.
	These two angle-based sequential pruning algorithms are summarized schematically in Fig.~\ref{figure1}.
	
	The main contributions of this work are threefold.
	First, we apply ma-QAOA to TFD-state preparation with SYK models and obtain high-fidelity TFD states even at low temperature.
	Second, we show that SAP-ma-QAOA retains high TFD fidelity with shallow circuit depth, while OSAP-ma-QAOA can further improve the fidelity.
	Third, we show that SAP- and OSAP-ma-QAOA remain effective with an energy-based objective, while combined batch--sequential pruning reduces the optimization cost.
	
	We substantiate these results by applying ma-QAOA and the proposed pruning algorithms to the Gaussian and binary $N=8$ SYK models in both dense and sparse cases, and to the binary sparse $N=10$ SYK model at $\beta=0.1$, $1$, and $10$.
	We quantify the performance of the algorithms by numerically computing the infidelity with the exact TFD state and the required circuit resources at each pruning step.
	We compare the infidelity trajectories of the sequential angle-pruning algorithms with those of the batch-pruning and random-pruning algorithms.
	We further verify the algorithms using the energy expectation value as an alternative objective function and combine sequential and batch pruning to reduce the optimization cost.
	
	The rest of this paper is organized as follows.
	In Sec.~\ref{sec:syk}, we introduce the dense and sparse SYK models and define the corresponding TFD target states.
	In Sec.~\ref{sec:qaoa}, we describe the QAOA-inspired TFD ansatz and its multi-angle generalization.
	In Sec.~\ref{sec:pruning}, we introduce SAP-ma-QAOA and OSAP-ma-QAOA, describe the comparison algorithms, and present the numerical results.
	In Sec.~\ref{sec:hardware}, we examine the two algorithms with a different objective function and combine sequential and batch pruning for hardware-efficient implementation.
	We conclude with a discussion of the results and possible extensions in Sec.~\ref{sec:discussion}.

	\section{\label{sec:syk}SYK models and thermofield double state}
	For our analysis, we consider four SYK models: (i) Gaussian dense, (ii) Gaussian sparse, (iii) binary dense, and (iv) binary sparse SYK models.
	In the following, we briefly describe these models and the corresponding TFD target state.
	
	\subsection{\label{sec:tfd_syk}SYK models}
	We consider a doubled SYK system consisting of a left and a right system, denoted as $L$ and $R$, respectively.
	Each side contains $N$ Majorana fermions, denoted by $\psi^{j}_{L}$ and $\psi^{j}_{R}$, with the Clifford algebra $\{\psi^{i}_{a},\psi^{j}_{b}\}=\delta_{a b}\delta^{i j}$ where $a,b=L,R$ and $i,j=1,\ldots,N$.
	For even $q$, the Gaussian dense SYK Hamiltonian on each side is
	\begin{equation}
		\label{eq:syk}
		H_{a}=i^{q/2}\sum_{1\leq j_{1}<\cdots<j_{q}\leq N}J_{j_{1}\cdots j_{q}}\,\psi^{j_{1}}_{a}\cdots\psi^{j_{q}}_{a},
	\end{equation}
	where the random couplings have
	\begin{equation}
		\overline{J_{j_{1}\cdots j_{q}}} = 0,
		\qquad
		\overline{J_{j_{1}\cdots j_{q}}^{2}} = \dfrac{J^{2}(q-1)!}{N^{q-1}}.
	\end{equation}
	Throughout this paper, we set $J=\sqrt{2}$.
	The two copies are combined into the two-sided Hamiltonian $H_{L}+H_{R}$.
	We also consider the left--right interaction Hamiltonian
	\begin{equation}
		\label{eq:mixer}
		H_{\rm int}=i\sum_{j=1}^{N}\psi^{j}_{L}\psi^{j}_{R}.
	\end{equation}
	Thus, the total Hamiltonian is given by
	\begin{equation}
		\label{eq:H_tot}
		H_{\rm tot}=H_{L}+H_{R}+gH_{\rm int},
	\end{equation}
	where $g$ is a coupling constant.
	
	For quantum-computer implementation, we adopt the Jordan--Wigner (JW) transformation
	\begin{equation}
		\psi_{L}^{i} = \dfrac{1}{\sqrt{2}}Z^{i - 1}X I^{N - i},
		\qquad
		\psi_{R}^{i} = \dfrac{1}{\sqrt{2}}Z^{i - 1}Y I^{N - i},
	\end{equation}
	where the products and powers are understood as tensor products.
	This transformation ensures that the Majorana operators satisfy the Clifford algebra and that $H_{\rm int}$ is Hermitian.
	
	\subsection{Sparse SYK models}
	We briefly review two classes of sparse SYK models.
	
	\smallskip
	\textit{1. Gaussian sparse SYK.}---
	The Gaussian sparse SYK Hamiltonian is given by
	\begin{align}
		\label{eq:sparse_1}
		H_{a} = i^{q/2}\sum_{1 \leq j_{1} < \cdots < j_{q} \leq N}\mathcal{J}_{j_{1}\cdots j_{q}}\psi_{a}^{j_{1}}\cdots\psi_{a}^{j_{q}},\\
		\label{eq:J}
		\mathcal{J}_{j_{1}\cdots j_{q}} = J_{j_{1}\cdots j_{q}}x_{j_{1}\cdots j_{q}},
	\end{align}
	where $x_{j_{1}\cdots j_{q}}\in\{0,1\}$ specifies whether the corresponding interaction term is retained.
	In the sparse ensemble, the nonzero Gaussian couplings have
	\begin{equation}
		\label{eq:sparse_syk_variance}
		\overline{J_{j_{1}\cdots j_{q}}} = 0,
		\qquad
		\overline{J_{j_{1}\cdots j_{q}}^{2}} = \frac{J^{2}(q-1)!}{p_{s}N^{q-1}},
	\end{equation}
	where $p_{s}$ denotes the retention probability, such that $x_{j_{1}\cdots j_{q}}=1$ with probability $p_{s}$.
	In our finite-$N$ numerics, we fix the number of retained interaction terms to $K=p_{s}\binom{N}{q}$ for a given $p_{s}$, where $p_{s}=1$ corresponds to the dense case.
	
	\smallskip
	\textit{2. Binary sparse SYK.}---
	Another sparsification scheme is the binary-coupling sparse SYK model~\cite{PhysRevB.107.L081103}.
	This model takes the coefficient in~\eqref{eq:sparse_1} in the form
	\begin{equation}
		\mathcal{J}_{j_{1}\cdots j_{q}} = x_{j_{1}\cdots j_{q}}\,\eta_{j_{1}\cdots j_{q}}\,\dfrac{J}{\sqrt{K}},
		\quad
		\eta_{j_{1}\cdots j_{q}}\in\{+1,-1\},
	\end{equation}
	with the two signs chosen with equal probability, while the retained interaction terms are chosen uniformly at random.
	All retained coefficients therefore have the same magnitude $J/\sqrt{K}$.
	
	It is known that the binary sparse SYK model retains quantum-chaotic behavior under sparsification more robustly than the Gaussian sparse SYK model~\cite{PhysRevB.107.L081103}.
	The binary sparse SYK model has been used for hardware-feasible quantum simulation for this reason~\cite{Byun:2026ewk,Bang:2026eof}. 
	For comparison, we also consider the binary dense SYK model with $p_{s} = 1$.
	For all four SYK models considered above, we focus on the $q=4$ case throughout this paper.
	
	\subsection{Thermofield double state}
	The TFD target state with a given SYK Hamiltonian at an inverse temperature $\beta = 1/T$ is prepared from a normalized maximally entangled state $\ket{I}$ by imaginary-time evolution,
	\begin{equation}
		\label{eq:tfd_target}
		\ket{\mathrm{TFD}(\beta)} = \dfrac{2^{N/2}}{\sqrt{Z_{\beta}}} \, e^{-\frac{\beta}{4}(H_{L} + H_{R})}\ket{I},
	\end{equation}
	where $Z_{\beta} = \Tr e^{-\beta H_{L}} = \Tr e^{-\beta H_{R}}$ in the JW transformation used here. 
	The maximally entangled state satisfies $(\psi_{L}^{j} + i\psi_{R}^{j})\ket{I} = 0 \, \forall j$ and is the ground state of $H_{\rm int}$.
	This is also equivalent to the infinite-temperature ($\beta=0$) limit of \eqref{eq:tfd_target}, and imaginary-time evolution under $H_{L} + H_{R}$ lowers $T$ to the target value. 
	We use \eqref{eq:tfd_target} as the variational target state throughout this work.
	
	\section{QAOA-motivated ansatz for TFD-state preparation}
	\label{sec:qaoa}
	
	\subsection{Quantum approximate optimization algorithm}
	We briefly review the QAOA-motivated circuit ansatz applied to SYK models.
	The QAOA constructs a variational state by alternating two noncommuting time evolutions, one generated by a cost Hamiltonian $H_{\rm cost}$ and the other by a mixer Hamiltonian $H_{\rm mix}$~\cite{Farhi2014qaoa,Hadfield2019}.
	Starting from a simple reference state, denoted as $\ket{\psi_{0}}$, a $p$-layer QAOA circuit is written as
	\begin{equation}
		\begin{aligned}
			\label{eq:qaoa_state}
			\ket{\psi_p(\boldsymbol{\gamma},\boldsymbol{\alpha})}
			&=
			\prod_{\ell=1}^{p}
			U_{\ell}(\gamma_{\ell},\alpha_{\ell})
			\ket{\psi_0},\\
			U_{\ell}(\gamma_{\ell},\alpha_{\ell})
			&=
			e^{-i\alpha_{\ell}H_{\rm mix}}
			e^{-i\gamma_{\ell}H_{\rm cost}},
		\end{aligned}
	\end{equation}
	where $\boldsymbol{\gamma}=(\gamma_1,\ldots,\gamma_p)$ and $\boldsymbol{\alpha}=(\alpha_1,\ldots,\alpha_p)$ denote the cost and mixer angles, respectively, and $\ell$ denotes the layer index.
	
	In our setting, the doubled SYK Hamiltonian serves as the cost Hamiltonian and the left--right coupling as the mixer Hamiltonian, so we identify $H_{\rm cost} = H_{L} + H_{R}$ and $H_{\rm mix} = H_{\rm int}$.
	The reference state is chosen as the maximally entangled state, such that $\ket{\psi_{0}} = \ket{I}$, so the final variational state becomes
	\begin{equation}
		\label{eq:qaoa_tfd_state}
		\ket{\psi_p(\boldsymbol{\gamma},\boldsymbol{\alpha})} = \prod_{\ell=1}^{p}e^{-i\alpha_{\ell}H_{\rm int}}e^{-i\gamma_{\ell}(H_{L} + H_{R})}\ket{I}.
	\end{equation}
	After the JW transformation, $H_{\rm cost}$ and $H_{\rm mix}$ become sums of Pauli strings,
	\begin{equation}
		H_{\rm cost}=\sum_{\nu = 1}^{M_C} c_{\nu} P_{\nu}^{C},
		\qquad
		H_{\rm mix}=\sum_{\mu=1}^{M_M} m_\mu P^M_\mu,
	\end{equation}
	where $P_{\nu}^{C}$ and $P_{\mu}^{M}$ are Pauli strings acting on the qubit Hilbert space, $c_{\nu}$ and $m_{\mu}$ are constants, and $M_C$ and $M_M$ denote the numbers of Pauli strings in $H_{\rm cost}$ and $H_{\rm mix}$, respectively.
	In our ma-QAOA setting, SYK models with system size $N$ generally give
	\begin{equation}
		M_{C} = 2K,
		\qquad
		M_{M} = N.
	\end{equation}
	Therefore, the time evolutions in QAOA can be decomposed into products of individual Pauli-string evolutions in the schematic form of $e^{-i\theta P}$ using the usual Trotterization.
	% Trotterization
	
	The variational parameters $\{\boldsymbol{\gamma}, \boldsymbol{\alpha}\}$ are optimized to minimize a cost function $\mathcal{L}(\boldsymbol{\gamma}, \boldsymbol{\alpha})$ at a desired $\beta$.
	We consider two choices for $\mathcal{L}$.
	The first is the infidelity with respect to the target TFD state, where the fidelity is defined by
	\begin{equation}
		\label{eq:qaoa_fidelity}
		F(\boldsymbol{\gamma},\boldsymbol{\alpha})
		=
		\left|\innerproduct{\mathrm{TFD}(\beta)}{\psi_p(\boldsymbol{\gamma},\boldsymbol{\alpha})}\right|^{2}.
	\end{equation}
	We maximize this quantity, or equivalently minimize the infidelity
	\begin{equation}
		\label{eq:qaoa_loss}
		\mathcal{L}(\boldsymbol{\gamma},\boldsymbol{\alpha})
		=
		1-F(\boldsymbol{\gamma},\boldsymbol{\alpha}).
	\end{equation}
	Alternatively, one can adopt a VQE approach; the TFD state can approximately be obtained from the ground state of $H_{\rm tot}$ in \eqref{eq:H_tot}, denoted as $\ket{\mathrm{GS}(g)}$.
	In the present finite-$N$ analysis, we choose $g$ such that $\abs{\innerproduct{\mathrm{GS}(g)}{\mathrm{TFD}(\beta)}}^{2} \approx 1$.
	Thus, the approximate TFD state can be obtained by minimizing the energy cost function\footnote{
		Variational approaches have also been developed to prepare Gibbs states directly by minimizing the free energy~\cite{PhysRevApplied.16.054035,Guo_2023,Araz:2024xkw,PhysRevA.110.012445,Selisko_2024,Robertson2026}.}
	\begin{equation}
		\mathcal{L}(\boldsymbol{\gamma}, \boldsymbol{\alpha}) = \bra{\psi_{p}(\boldsymbol{\gamma}, \boldsymbol{\alpha})}H_{\rm tot}\ket{\psi_{p}(\boldsymbol{\gamma}, \boldsymbol{\alpha})},
	\end{equation}
	so that the full optimization is expected to give $\ket{\psi_{p}(\boldsymbol{\gamma}, \boldsymbol{\alpha})}\approx\ket{\mathrm{GS}(g)}$.
	While determining $g$ in this way still requires the exact TFD state, we use the energy objective to examine whether our pruning algorithms remain effective with a quantum-measurable objective.
	In coupled SYK models, however, the relation between $g$ and $\beta$ is known~\cite{Maldacena:2018lmt,Alet2021}, providing a practical way to estimate $g$ for a target $\beta$ in an energy-based implementation.
	Unless otherwise stated, we use the infidelity objective; the energy objective is used in Sec.~\ref{sec:hardware}.
	
	\subsection{Multi-angle QAOA}
	In the standard QAOA, all time evolutions generated by $H_{\rm cost}$ within a layer share a single angle $\gamma_{\ell}$, while all time evolutions generated by $H_{\rm mix}$ share another angle $\alpha_{\ell}$.
	Here we instead use a ma-QAOA ansatz~\cite{Herrman2022}, in which every time evolution carries an independent angle,
	\begin{equation}
		\label{eq:maqaoa}
		|\psi(\bm\theta)\rangle = \prod_{\ell=1}^{p}\left[\prod_{\mu=1}^{M_M}e^{-i\theta^M_{\ell,\mu}P^M_\mu}\prod_{\nu = 1}^{M_C}e^{-i\theta^C_{\ell,\nu}P^{C}_{\nu}}\right]\ket{I},
	\end{equation}
	for a given $p$, where $\bm\theta=\{\theta^{C}_{\ell,\nu},\theta^M_{\ell,\mu}\}$.
	We refer to the Pauli-string evolutions $e^{-i\theta^C_{\ell,\nu}P_{\nu}^{C}}$ and $e^{-i\theta^M_{\ell,\mu}P^M_\mu}$ as cost and mixer evolutions, respectively.
	
	\begin{figure}[tb]
		\centering
		\includegraphics[width=0.9\linewidth]{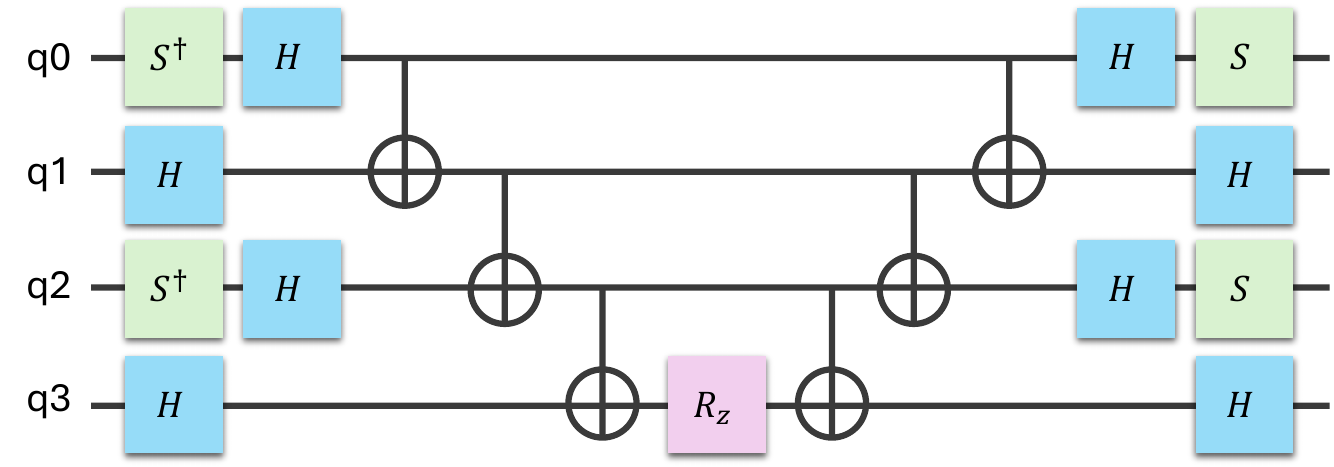}
		\caption{\label{figure2}
			Circuit realization of a single cost evolution for the cost term $\psi_L^{1}\psi_L^{2}\psi_L^{3}\psi_L^{4}$. The identity operators act on the other four qubits for $N = 8$ SYK models.
		}
	\end{figure}
	
	Each cost evolution can be decomposed into a ladder of CNOT gates around an $R_{z}$ gate, together with $H$, $S$, and $S^{\dagger}$ gates.
	This structure is common to every cost evolution; for $j_1<j_2<j_3<j_4$, the JW transformation maps a cost term to
	\begin{equation}
		\label{eq:cost_string}
		\psi_a^{j_1}\psi_a^{j_2}\psi_a^{j_3}\psi_a^{j_4} \sim A_{j_1}B_{j_2}A_{j_3}B_{j_4}\prod_{k=j_1+1}^{j_2-1}Z_{k}\prod_{k=j_3+1}^{j_4-1}Z_{k},
	\end{equation}
	where $(A,B)=(Y,X)$ and $(X,Y)$ for $a=L$ and $R$, respectively.
	Here the subscript denotes the qubit position and the identity acts on the remaining qubits.
	An evolution therefore acts on $w = j_{2} + j_{4} - j_{1} - j_{3} + 2$ qubits with $2(w-1)$ CNOT gates.
	After the basis change, the Pauli string becomes a product of $Z$ operators, so the phase depends only on the parity of the qubits involved.
	The CNOT chain accumulates this parity on the last qubit, where a single $R_z$ rotation applies the phase, and is then reversed.
	Since $w$ depends on the interaction indices, different cost evolutions can have different circuit depths and gate counts.
	Also, since the cost evolutions generally do not commute, their ordering is fixed for each disorder realization and kept unchanged throughout the optimization.
	Figure~\ref{figure2} shows a representative cost evolution from the cost term $\psi_L^{1}\psi_L^{2}\psi_L^{3}\psi_L^{4}$.
	
	On the other hand, $H_{\rm mix}$ becomes a sum of single-qubit operators under the JW transformation,
	\begin{equation}
		H_{\rm mix} = -\frac{1}{2}\sum_{j=1}^{N} Z_j .
		\label{eq:mixer_jw}
	\end{equation}
	Thus, each mixer evolution acts only on a single qubit.
	Since the cost evolutions require two-qubit gates in their circuit decomposition, we refer to each of them as a \textit{nonlocal block}, or simply \textit{block}, and we count only these blocks when referring to the circuit block count.
	Therefore, the initial ma-QAOA ansatz contains $pM_{C} = 2pK$ blocks.
	In addition, for a consistent comparison, we evaluate the circuit depth and gate counts after simplifying only the mixer evolutions, using the same gate set as in the ma-QAOA circuit, $\{\mathrm{CNOT}, R_{z}, H, S, S^{\dagger}\}$.
	
	Assigning an independent parameter to each evolution in ma-QAOA increases the number of variational parameters relative to the standard QAOA.
	For example, the dense $N = 8$ Hamiltonian produces $p(M_C+M_M) = 148p$ independent variational parameters before any simplification.
	Despite this large parameter space, optimizing each evolution allows the circuit to reach the target TFD state with high fidelity.\footnote{One can reduce the optimization cost of ma-QAOA through parameter freezing~\cite{Jang:2026} and parameter reduction~\cite{Shi9996634,Kim:2026pnn}.}
	As in \eqref{eq:qaoa_fidelity} and \eqref{eq:qaoa_loss}, the objective function is the infidelity with the exact TFD state,
	\begin{equation}
		\mathcal{L}(\bm\theta)=1-F(\bm\theta), \qquad F(\bm\theta)=\left|\langle \mathrm{TFD}(\beta)|\psi(\bm\theta)\rangle\right|^{2}.
		\label{eq:fidelity}
	\end{equation}
	
	\begin{table}[t]
		\centering
		\footnotesize
		\begin{ruledtabular}
			\begin{tabular}{llcc}
				Model & $\beta$
				& QAOA $\mathbb{E}_{J}[F]$
				& ma-QAOA $\mathbb{E}_{J}[F]$ \\
				\hline
				\multirow{3}{*}{Gaussian dense}
				& $0.1$ & $1.000000\,(0)$ & $1.000000\,(0)$ \\
				& $1$   & $0.999999\,(1\times10^{-6})$ & $1.000000\,(0)$ \\
				& $10$  & $0.920725\,(0.029025)$ & $1.000000\,(0)$ \\
				\hline
				\multirow{3}{*}{Gaussian sparse}
				& $0.1$ & $1.000000\,(0)$ & $1.000000\,(0)$ \\
				& $1$   & $0.999999\,(1\times10^{-6})$ & $1.000000\,(0)$ \\
				& $10$  & $0.922651\,(0.030152)$ & $0.999075\,(0.000328)$ \\
				\hline
				\multirow{3}{*}{Binary dense}
				& $0.1$ & $1.000000\,(0)$ & $1.000000\,(0)$ \\
				& $1$   & $0.999997\,(2\times10^{-6})$ & $1.000000\,(0)$ \\
				& $10$  & $0.886236\,(0.030025)$ & $1.000000\,(0)$ \\
				\hline
				\multirow{3}{*}{Binary sparse}
				& $0.1$ & $1.000000\,(0)$ & $1.000000\,(0)$ \\
				& $1$   & $0.999998\,(1\times10^{-6})$ & $0.999999\,(1\times10^{-6})$ \\
				& $10$  & $0.901556\,(0.032279)$ & $0.987927\,(0.008990)$ \\
			\end{tabular}
		\end{ruledtabular}
		\caption{\label{table1}
			Comparison of disorder-averaged fidelities $\mathbb{E}_{J}[F]$ between standard QAOA and ma-QAOA for the four $N=8$ SYK models, averaged over 20 disorder realizations.
			Each entry reports the mean fidelity with the standard deviation in parentheses.
			For the sparse models, Gaussian sparse uses $K=32$ and binary sparse uses $K=10$.
			All quantum circuits use $p=3$ layers.
		}
	\end{table}
	
	We first numerically compare ma-QAOA with standard QAOA.
	Table~\ref{table1} shows disorder-averaged fidelities $\mathbb{E}_{J}[F]$ and their standard deviations $\sigma_{J}[F]$ for QAOA and ma-QAOA for the four SYK models at $\beta=0.1$, $1$, and $10$.
	As shown in this table, ma-QAOA achieves high fidelity with the TFD state over the temperatures considered here.
	In particular, ma-QAOA outperforms QAOA at low temperature.
	This result shows that ma-QAOA provides a more effective ansatz for constructing the full variational circuit for the TFD state.
	
	\section{Sequential angle-pruning ma-QAOA}
	\label{sec:pruning}
	
	The ma-QAOA ansatz can contain a large number of nonlocal blocks, particularly for Hamiltonians with many interaction terms such as the SYK model.
	To reduce the resulting circuit cost while retaining high fidelity with the target TFD state, we consider pruning these blocks based on their optimized angle magnitudes.

	\subsection{Sequential pruning algorithms}
	We introduce two pruning algorithms for the ma-QAOA circuit, referred to as SAP-ma-QAOA and OSAP-ma-QAOA.
	We first introduce SAP-ma-QAOA.
	
	\smallskip	
	\textit{1. SAP-ma-QAOA.}---
	This algorithm sequentially removes the nonlocal block with the smallest $|\theta|$ and reoptimizes the remaining parameters after each removal.
	
	The pruning procedure starts from the optimized initial ma-QAOA ansatz.
	Let $\mathcal{B}=\{b_j\}$ be the set of nonlocal blocks in the optimized ansatz, where $\theta_j\in\{\theta^{C}_{\ell,\nu}\}\subset\bm\theta$ denotes the optimized angle associated with block $b_j$.
	The goal is to reduce $\mathcal{B}$ to a prescribed target size $\mathcal{S}$ while keeping the cost function $\mathcal{L}$ as low as possible.
	
	To motivate the small-angle pruning criterion, we consider the change in the cost function when a block $b_{j}$ is removed from the optimized circuit.
	Around the optimized parameters $\bm{\theta}^{\ast}$, the first-order variation is assumed to be negligible, so the cost function is locally given by
	\begin{equation}
		\mathcal{L}(\bm{\theta}^{\ast}+\delta\bm{\theta}) \simeq \mathcal{L}(\bm{\theta}^{\ast})+\frac{1}{2}\sum_{i,k}\mathcal{H}_{ik}\delta\theta_i\delta\theta_k,
	\end{equation}
	where $\mathcal{H}_{ik} = \left.{\partial^{2}\mathcal{L}}/{\partial\theta_{i}\partial\theta_{k}}\right|_{\bm{\theta}^{\ast}}$ is the Hessian.
	Removing the block associated with $\theta_j^{\ast}$ corresponds to setting $\theta_j^{\ast}$ to zero while keeping the other parameters $\bm{\theta}_{\bar j}^{\ast} = \bm{\theta}^{\ast}\setminus\{\theta_{j}^{\ast}\}$ fixed.
	Therefore, the cost change immediately after the removal is approximately
	\begin{equation}
		\Delta \mathcal{L}_{j}^{\rm pre} := \mathcal{L}(\bm{\theta}^{\ast}+\delta\bm{\theta}) - \mathcal{L}(\bm{\theta}^{\ast}) \simeq\frac{1}{2}\mathcal{H}_{jj}\left(\theta_j^{\ast}\right)^2.
	\end{equation}
	When the variation of $\mathcal{H}_{jj}$ among the blocks is not large, this relation indicates that removing a block with a smaller $|\theta_j^{\ast}|$ produces a smaller perturbation $\Delta \mathcal{L}_{j}^{\rm pre}$ to the optimized cost function.
	
	After removing $b_{j}$, the remaining parameters $\bm{\theta}_{\bar j}^{\ast}$ are then reoptimized.
	Denoting the remaining parameters during reoptimization by $\bm{\theta}_{\bar j}$, the reoptimized cost is defined as
	\begin{equation}
		\label{eq:L_post}
		\mathcal{L}_{j}^{\rm post} := \min_{\bm{\theta}_{\bar j}}\mathcal{L}(0,\bm{\theta}_{\bar j})\leq \mathcal{L}_{j}^{\rm pre},
	\end{equation}
	where $\mathcal{L}_{j}^{\rm pre} := \mathcal{L}(0,\bm{\theta}_{\bar j}^{\ast})$ denotes the cost immediately after removing $b_{j}$.
	Thus, SAP-ma-QAOA first selects a block whose removal is expected to produce a small perturbation to the optimized cost and then reoptimizes the remaining parameters to further reduce the cost.
	
	\begin{algorithm}[tb]
		\caption{SAP-ma-QAOA: sequential small-angle pruning with reoptimization}
		\label{algorithm1}
		\KwIn{$\bm{\theta}$, $\mathcal{B}$, $\mathcal{L}$, $\mathcal{S}$}
		\KwOut{$\bm{\theta}_{\rm pr}$, $\mathcal{B}_{\rm pr}$}
		
		Initialize $\mathcal{B}_{\rm act}\gets\mathcal{B}$ and $\bm{\theta}_{\rm act}\gets\bm{\theta}$\;
		
		\While{$\mathcal{N}(\mathcal{B}_{\rm act})>\mathcal{S}$}{
			Select the smallest-angle block,
			\[
			j^{\star}\gets\argmin_{j:\,b_{j}\in\mathcal{B}_{\rm act}}|\theta_{j}|.
			\]
			
			Remove the selected block and its angle,
			\[
			\mathcal{B}_{\rm act}\gets\mathcal{B}_{\rm act}\setminus\{b_{j^{\star}}\}, \qquad
			\bm{\theta}_{\rm act}\gets\bm{\theta}_{\rm act}\setminus\{\theta_{j^{\star}}\}.
			\]
			
			Reoptimize the remaining parameters,
			\[
			\bm{\theta}_{\rm act}\gets\argmin_{\bm{\theta}_{\rm act}}\mathcal{L}(\bm{\theta}_{\rm act}).
			\]
		}
		
		Set $\mathcal{B}_{\rm pr}\gets\mathcal{B}_{\rm act}$ and $\bm{\theta}_{\rm pr}\gets\bm{\theta}_{\rm act}$\;
		
		\Return{$\bm{\theta}_{\rm pr}$, $\mathcal{B}_{\rm pr}$}
	\end{algorithm}
	
	In SAP-ma-QAOA, the nonlocal blocks are removed sequentially according to their angle magnitudes.
	At each pruning round, let $\mathcal{B}_{\rm act}\subseteq\mathcal{B}$ denote the set of active nonlocal blocks, and let $\bm{\theta}_{\rm act}\subseteq\bm{\theta}$ denote the active variational parameters, consisting of $\{\theta^{M}_{\ell,\mu}\}$ and $\{\theta_j:b_j\in\mathcal{B}_{\rm act}\}$.
	We denote by $\mathcal{N}(\mathcal{A})$ the number of blocks in a block set $\mathcal{A}$.
	In the initial ma-QAOA, we have $\mathcal{B}_{\rm act}=\mathcal{B}$ and $\bm{\theta}_{\rm act}=\bm{\theta}$.
	We select the block $b_{j^\star}\in\mathcal{B}_{\rm act}$ with the smallest angle magnitude,
	\begin{equation}
		\label{eq:sap_prune_rule}
		j^\star=\argmin_{j:\,b_j\in\mathcal{B}_{\rm act}}|\theta_j|.
	\end{equation}
	We then update
	\begin{equation}
		\label{eq:sap_removal}
		\mathcal{B}_{\rm act}\leftarrow\mathcal{B}_{\rm act}\setminus\{b_{j^\star}\}, \qquad
		\bm{\theta}_{\rm act}\leftarrow\bm{\theta}_{\rm act}\setminus\{\theta_{j^\star}\}.
	\end{equation}
	The remaining active variational parameters are reoptimized as
	\begin{equation}
		\label{eq:sap_reoptimization}
		\bm{\theta}_{\rm act}\leftarrow\argmin_{\bm{\theta}_{\rm act}}\mathcal{L}(\bm{\theta}_{\rm act}),
	\end{equation}
	before the next pruning round.
	For each reoptimization, the search range for every parameter in $\bm{\theta}_{\rm act}$ is set to its current value plus $[-\pi,\pi)$, with the current value used as the initial point.
	The next block is selected according to the reoptimized angles.
	The procedure is repeated until $\mathcal{N}(\mathcal{B}_{\rm act})=\mathcal{S}$.
	The final active block set and variational parameters are denoted by $\mathcal{B}_{\rm pr}$ and $\bm{\theta}_{\rm pr}$, respectively.
	We summarize the whole procedure in Algorithm~\ref{algorithm1}.
	
	The angle magnitude, however, only estimates the cost perturbation before reoptimization.
	Although reoptimization can only reduce the cost from $\mathcal{L}_{j}^{\rm pre}$, as shown in \eqref{eq:L_post}, the amount of this reduction can differ depending on which block is removed.
	Therefore, removing $b_{j^{\star}}$ does not necessarily yield the lowest post-reoptimization cost.
	
	To account for this effect, we introduce OSAP-ma-QAOA.
	
	\smallskip
	\textit{2. OSAP-ma-QAOA.}---
	This algorithm tentatively removes candidate nonlocal blocks with small angle magnitudes, reoptimizes the remaining parameters for each candidate, and selects the candidate that yields the lowest post-reoptimization cost.
	
	\begin{algorithm}[tb]
		\caption{OSAP-ma-QAOA: objective-aware sequential small-angle pruning}
		\label{algorithm2}
		\KwIn{$\bm{\theta}$, $\mathcal{B}$, $\mathcal{L}$, $\mathcal{S}$, $N_{\mathcal C}$}
		\KwOut{$\bm{\theta}_{\rm pr}$, $\mathcal{B}_{\rm pr}$}
		
		Initialize $\mathcal{B}_{\rm act}\gets\mathcal{B}$ and $\bm{\theta}_{\rm act}\gets\bm{\theta}$\;
		
		\While{$\mathcal{N}(\mathcal{B}_{\rm act})>\mathcal{S}$}{
			Form the candidate pool $\mathcal{C}\subseteq\mathcal{B}_{\rm act}$ with
			\[
			\mathcal{N}(\mathcal{C})=\min\!\left(N_{\mathcal C},\mathcal{N}(\mathcal{B}_{\rm act})\right),
			\]
			consisting of the blocks with the smallest angle magnitudes\;
			
			\ForEach{$b_{j}\in\mathcal{C}$}{
				$\bm{\theta}_{\rm act}^{(j)}
				\gets
				\argmin_{\bm{\theta}_{\rm act}\setminus\{\theta_{j}\}}
				\mathcal{L}\!\left(\bm{\theta}_{\rm act}\setminus\{\theta_{j}\}\right)$\;
				
				$\mathcal{L}_{j}^{\rm post}
				\gets
				\mathcal{L}\big(\bm{\theta}_{\rm act}^{(j)}\big)$\;
			}
			
			Select the candidate with the lowest post-reoptimization cost,
			\[
			j^{\star}\gets\argmin_{j:\,b_{j}\in\mathcal{C}}\mathcal{L}_{j}^{\rm post}.
			\]
			
			Update the active block set and parameters,
			\[
			\mathcal{B}_{\rm act}\gets\mathcal{B}_{\rm act}\setminus\{b_{j^{\star}}\}, \qquad
			\bm{\theta}_{\rm act}\gets\bm{\theta}_{\rm act}^{(j^{\star})}.
			\]
		}
		
		Set $\mathcal{B}_{\rm pr}\gets\mathcal{B}_{\rm act}$ and $\bm{\theta}_{\rm pr}\gets\bm{\theta}_{\rm act}$\;
		
		\Return{$\bm{\theta}_{\rm pr}$, $\mathcal{B}_{\rm pr}$}
	\end{algorithm}
	
	At each pruning round, we form a candidate pool $\mathcal{C}\subseteq\mathcal{B}_{\rm act}$ consisting of the blocks $b_j\in\mathcal{B}_{\rm act}$ with the smallest $|\theta_j|$.
	We prescribe the candidate-pool size as $N_{\mathcal C}$, while the actual number of blocks in $\mathcal{C}$ is given by
	\begin{equation}
		\label{eq:osap_candidate_size}
		\mathcal{N}(\mathcal{C}) = \min\!\left(N_{\mathcal C},\mathcal{N}(\mathcal{B}_{\rm act})\right).
	\end{equation}
	Thus, if $\mathcal{N}(\mathcal{B}_{\rm act}) \leq N_{\mathcal C}$, we have $\mathcal{C} = \mathcal{B}_{\rm act}$.
	For each candidate block $b_j\in\mathcal{C}$, we tentatively apply the removal and reoptimization steps in \eqref{eq:sap_removal} and \eqref{eq:sap_reoptimization}, respectively, with $j^\star$ replaced by $j$.
	This gives the reoptimized parameters $\bm{\theta}_{\rm act}^{(j)}$ and the corresponding post-reoptimization cost $\mathcal{L}_{j}^{\rm post}=\mathcal{L}(\bm{\theta}_{\rm act}^{(j)})$.
	We then select the candidate that yields the lowest post-reoptimization cost,
	\begin{equation}
		\label{eq:osap_prune_rule}
		j^\star=\argmin_{j:\,b_j\in\mathcal{C}}\mathcal{L}_{j}^{\rm post},
	\end{equation}
	and update $\mathcal{B}_{\rm act}\leftarrow\mathcal{B}_{\rm act}\setminus\{b_{j^\star}\}$ and $\bm{\theta}_{\rm act}\leftarrow\bm{\theta}_{\rm act}^{(j^\star)}$.
	The procedure is then repeated, as in SAP-ma-QAOA, until the target size $\mathcal{S}$ is reached.
	We summarize the whole procedure in Algorithm~\ref{algorithm2}.
	
	For the comparisons, we consider two more algorithms based on batch angle pruning (BAP), in which the blocks are removed at once according to the angles of the optimized initial ma-QAOA circuit.
	
	\smallskip
	\textit{3. BAP-ma-QAOA.}---
	This algorithm keeps the $\mathcal{S}$ nonlocal blocks with the largest $|\theta|$ in a single batch and then reoptimizes the retained parameters.
	
	\smallskip
	\textit{4. BAP$_{0}$-ma-QAOA.}---
	This algorithm uses the same selection rule as in BAP-ma-QAOA but evaluates the resulting circuit without reoptimization.
	
	\begin{table}[t]
		\centering
		\begin{tabular*}{1.0\linewidth}{@{\extracolsep{\fill}}lccc}
			\hline
			Method & Pruning & Objective & Reoptimization \\
			\hline
			SAP  & Sequential & No  & Once per round \\
			OSAP & Sequential & Yes & $\mathcal{N}(\mathcal{C})$ per round \\
			BAP  & One-shot   & No  & Once in total \\
			BAP$_{0}$ & One-shot & No & No \\
			\hline
		\end{tabular*}
		\caption{\label{table2}
			Comparison of the four pruning algorithms applied to the ma-QAOA ansatz.
			The Objective column indicates whether the post-reoptimization cost is used for block selection, and $\mathcal{N}(\mathcal{C})$ denotes the number of blocks in the candidate pool $\mathcal{C}$.
		}
	\end{table}
	
	Comparing BAP$_{0}$ with BAP shows the effect of reoptimization, while comparing BAP with SAP shows the effect of sequential pruning with intermediate reoptimization relative to one-shot pruning.
	Comparing SAP with OSAP tests whether the objective-aware choice of the removed block further improves the fidelity-resource tradeoff.
	The four algorithms are summarized in Table~\ref{table2}.
	
	\begin{figure*}[tb]
		\centering
		\includegraphics[width=\linewidth]{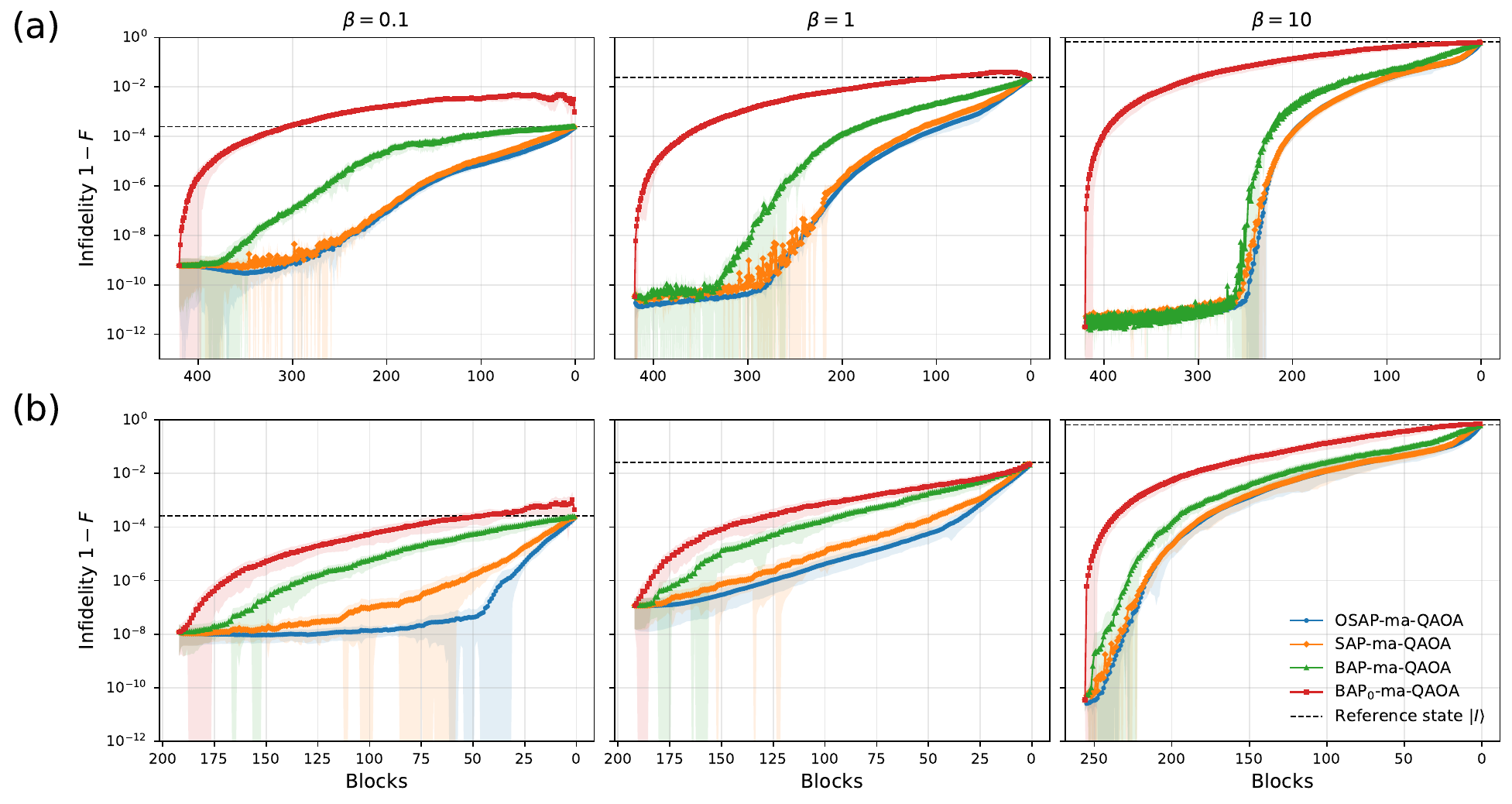}
		\caption{\label{figure3}
			Infidelity trajectories of the four pruning algorithms for the Gaussian $N=8$ SYK models at $\beta=0.1$, $1$, and $10$:
			(a) dense SYK and (b) sparse SYK with $K=32$.
			The curves and shaded regions show $\mathbb{E}_{J}[1 - F]$ and $\sigma_{J}[1 - F]$, respectively, averaged over 20 disorder realizations.
			The black dashed lines indicate the disorder-averaged reference infidelity between the exact TFD state and the maximally entangled state $\ket{I}$.
		}
	\end{figure*}
	
	\begin{figure*}[tb]
		\centering
		\includegraphics[width=\linewidth]{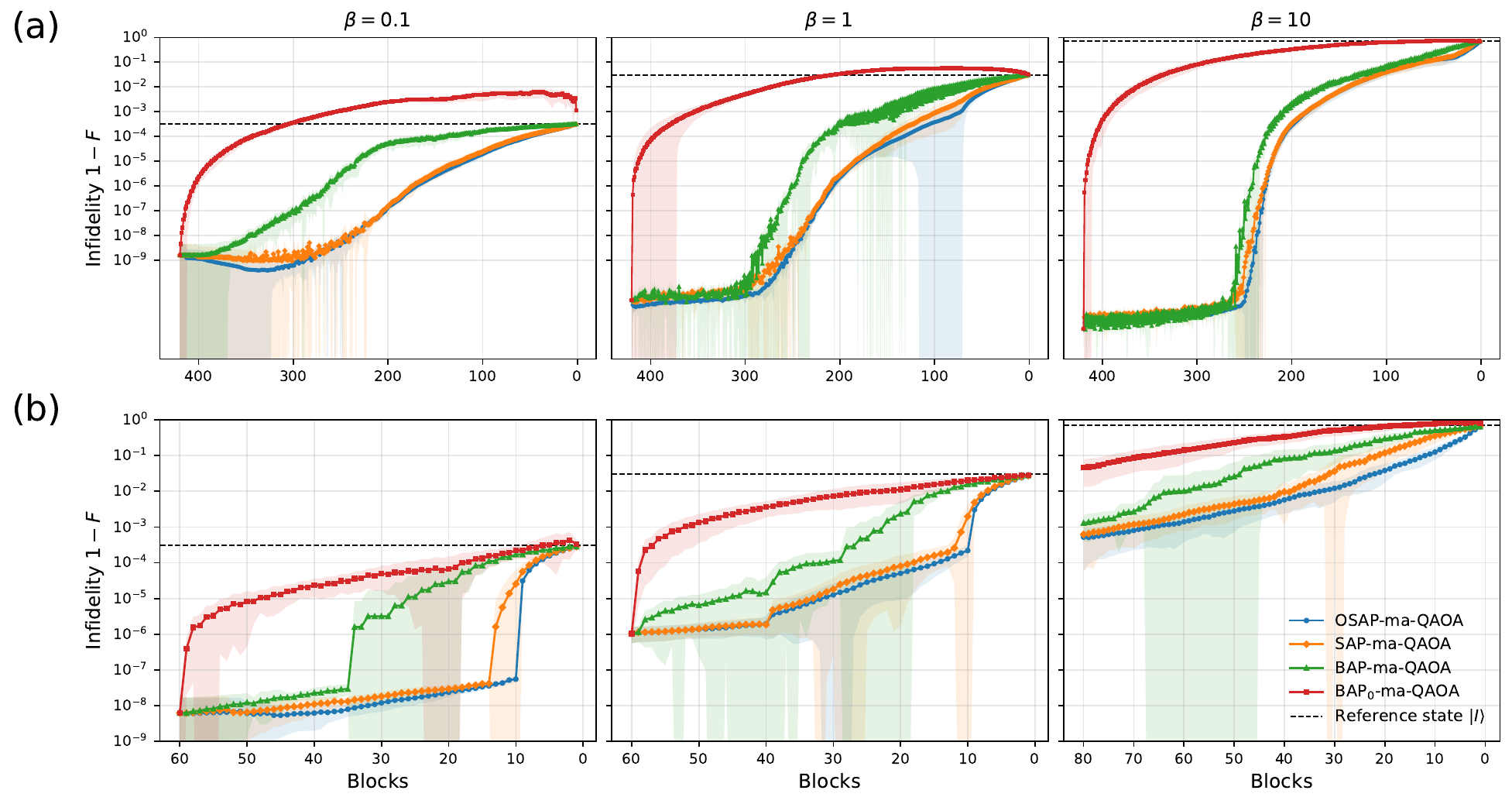}
		\caption{
			\label{figure4}
			Infidelity trajectories of the four pruning algorithms for the binary $N=8$ SYK models at $\beta=0.1$, $1$, and $10$:
			(a) dense SYK and (b) sparse SYK with $K=10$.
			The curves and shaded regions show $\mathbb{E}_{J}[1 - F]$ and $\sigma_{J}[1 - F]$, respectively, averaged over 20 disorder realizations.
			The black dashed lines indicate the disorder-averaged reference infidelity between the exact TFD state and the maximally entangled state $\ket{I}$.
		}
	\end{figure*}
	
	\subsection{\label{subsec:results}Performance of the pruning algorithms}
	
	We numerically compare SAP-, OSAP-, BAP-, and BAP$_0$-ma-QAOA for the four SYK models described in Sec.~\ref{sec:syk}.
	Unless otherwise stated, we consider $N=8$ SYK models at $\beta=0.1$, $1$, and $10$.
	We take $\mathcal{L}=1-F$ as the cost function and set $N_{\mathcal C}=10$.
	The four algorithms are compared at the same $\mathcal{S}$, while circuit depths and gate counts are reported separately.
	We show the trajectories of disorder-averaged infidelity $\mathbb{E}_{J}[1-F]$ and the corresponding standard deviation $\sigma_{J}[1 - F]$ over 20 disorder realizations for each SYK model.
	We plot them on a logarithmic scale to clearly distinguish the performance of the algorithms.
	For reference, the figures also show $\mathbb{E}_{J}[1-\abs{\innerproduct{\mathrm{TFD}(\beta)}{I}}^{2}]$, corresponding to the infidelity at $\mathcal{S}=0$ when all nonlocal blocks are removed.
	
	For each disorder realization, we start from $p=3$ and increase $p$ when necessary until the optimized initial ma-QAOA circuit reaches $F\geq99.99\%$ before pruning.
	At $\beta=0.1$ and $1$, the initial ma-QAOA circuit reaches $F\geq99.99\%$ for all disorder realizations of all four SYK models at $p=3$.
	At $\beta=10$, the Gaussian dense, Gaussian sparse, and binary dense SYK models reach $F\geq99.99\%$ for all disorder realizations at $p=3$, $4$, and $3$, respectively, while for the binary sparse SYK model the required $p$ varies from $4$ to $7$ across disorder realizations.
	Therefore, for the binary sparse SYK model at $\beta=10$, we plot the common infidelity trajectories from $\mathcal{S}=80$, corresponding to $p=4$.
	
	Throughout this work, the variational parameters are optimized using the L-BFGS-B method, with the objective-function gradients evaluated analytically.
	For the initial ma-QAOA optimization, we use five random initializations for each $p$, and the maximum number of iterations for each optimization run is set to 3000.
	Also, for each pruning reoptimization, the maximum number of iterations is set to 300.
	
	Figures~\ref{figure3} and~\ref{figure4} show $\mathbb{E}_{J}[1-F]$ for the Gaussian and binary $N=8$ SYK models, respectively, including both dense and sparse cases.
	As these figures show, SAP- and OSAP-ma-QAOA generally maintain lower infidelities than BAP- and BAP$_0$-ma-QAOA over the pruning trajectories at all temperatures considered here.
	Also, OSAP-ma-QAOA generally retains slightly lower infidelities than SAP-ma-QAOA over the trajectories.
	The same ordering generally holds for a single disorder realization, where the fidelities are compared at the same $\mathcal{S}$ at each $\beta$; see Appendix~\ref{sec:single_realization}.
	In OSAP-ma-QAOA, increasing $N_{\mathcal C}$ can further improve the performance at the expense of increased computational cost; the dependence on $N_{\mathcal C}$ is examined in Appendix~\ref{sec:pool_size}.
	
	\begin{table*}[t]
		\begin{ruledtabular}
			\begin{tabular}{lccccc}
				Method & Blocks & $\mathbb{E}_{J}[F]$ & Depth & $1q$ & $2q$ \\
				\hline
				SAP & 31\,(7) & 0.979152\,($4.399\times10^{-3}$) & 440\,(98) & 456\,(91) & 292\,(69) \\
				OSAP & 25\,(5) & 0.979750\,($1.774\times10^{-3}$) & 346\,(84) & 367\,(73) & 229\,(61) \\
				BAP & 46\,(11) & 0.982122\,($4.356\times10^{-3}$) & 639\,(147) & 648\,(143) & 426\,(103) \\
				BAP$_{0}$ & 90\,(14) & 0.979890\,($0.925\times10^{-3}$) & 1234\,(208) & 1217\,(192) & 827\,(150) \\
			\end{tabular}
		\end{ruledtabular}
		\caption{\label{table3}
			Comparison of the four ma-QAOA pruning algorithms for the binary sparse $N=8$ SYK model with $K=10$ at $\beta=10$ and fixed target fidelity $F=0.98$, averaged over 20 disorder realizations.
			For each realization and each algorithm, we select $\mathcal{S}$ for which the fidelity is closest to $F=0.98$ and report the mean values with the standard deviations in parentheses.
			Columns list the mean number of blocks, disorder-averaged fidelity $\mathbb{E}_{J}[F]$, mean circuit depth, and single-qubit ($1q$) and two-qubit ($2q$) gate counts.
		}
	\end{table*}
	
	To further assess the circuit reduction, we consider a fixed-fidelity comparison for the binary sparse $N=8$ SYK model with $K=10$ at $\beta=10$.
	For each disorder realization and each algorithm, we identify $\mathcal{S}$ for which the fidelity is closest to $F=0.98$, and report the disorder-averaged circuit resources in Table~\ref{table3}.
	As this table shows, OSAP-ma-QAOA achieves $F\approx0.98$ with the fewest blocks and the lowest circuit depth and gate counts.
	SAP-ma-QAOA requires more resources, while the two BAP algorithms require substantially larger circuits.
	Thus, the advantage of sequential angle pruning with reoptimization also translates into a more compact circuit at fixed target fidelity.
	
	\begin{figure*}[tb]
		\centering
		\includegraphics[width=\linewidth]{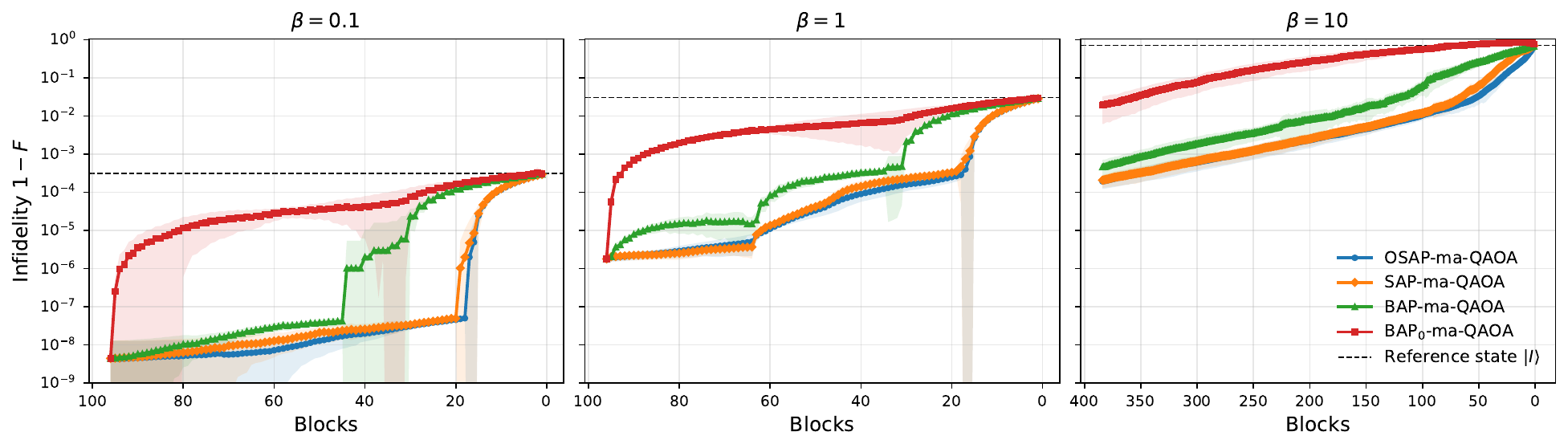}
		\caption{
			\label{figure5}
			Infidelity trajectories of the four pruning algorithms for the binary sparse $N=10$ SYK model with $K=16$ at $\beta=0.1$, $1$, and $10$.
			The curves and shaded regions show $\mathbb{E}_{J}[1 - F]$ and $\sigma_{J}[1 - F]$, respectively, averaged over 20 disorder realizations.
			The black dashed lines indicate the disorder-averaged reference infidelity between the exact TFD state and the maximally entangled state $\ket{I}$.
		}
	\end{figure*}
	
	\begin{table*}[t]
		\begin{ruledtabular}
			\begin{tabular}{clccccc}
				$\beta$ & Method & Blocks & $\mathbb{E}_{J}[F]$ & Depth & $1q$ & $2q$ \\
				\hline
				\multirow{4}{*}{$0.1$}
				& SAP  & 1 & 0.999707\,($<\!10^{-6}$) & 16\,(4) & 33\,(0) & 10\,(4) \\
				& OSAP & 1 & 0.999707\,($<\!10^{-6}$) & 16\,(3) & 32\,(4) & 9\,(3) \\
				& BAP  & 1 & 0.999707\,($<\!10^{-6}$) & 17\,(3) & 32\,(2) & 11\,(3) \\
				& BAP$_{0}$ & 1 & 0.999693\,($4.1\times10^{-5}$) & 17\,(3) & 32\,(2) & 11\,(3) \\
				\hline
				\multirow{4}{*}{$1$}
				& SAP & 1 & 0.971592\,($3.83\times10^{-4}$) & 17\,(3) & 28\,(5) & 11\,(3) \\
				& OSAP & 1 & 0.971635\,($3.57\times10^{-4}$) & 16\,(3) & 28\,(5) & 10\,(3) \\
				& BAP & 1 & 0.971517\,($3.97\times10^{-4}$) & 16\,(3) & 32\,(2) & 9\,(3) \\
				& BAP$_{0}$ & 1 & 0.970756\,($5.07\times10^{-4}$) & 16\,(3) & 32\,(2) & 9\,(3) \\
				\hline
				\multirow{4}{*}{$10$}
				& SAP & 43 & 0.907769\,($3.6364\times10^{-2}$) & 657\,(38) & 688\,(18) & 454\,(41) \\
				& OSAP & 43 & 0.951707\,($2.3653\times10^{-2}$) & 663\,(44) & 685\,(14) & 459\,(44) \\
				& BAP & 43 & 0.714790\,($5.7844\times10^{-2}$) & 654\,(48) & 672\,(13) & 452\,(50) \\
				& BAP$_{0}$ & 43 & 0.217304\,($1.05935\times10^{-1}$) & 654\,(48) & 672\,(13) & 452\,(50) \\
			\end{tabular}
		\end{ruledtabular}
		\caption{\label{table4}
			Comparison of the four ma-QAOA pruning algorithms for the binary sparse $N=10$ SYK model with $K=16$ at the same target block count $\mathcal{S}$, averaged over 20 disorder realizations.
			For each realization and each algorithm, we evaluate the circuit at fixed $\mathcal{S}$ and report the mean values with the standard deviations in parentheses.
			We set $\mathcal{S}=1$ for $\beta=0.1$ and $1$, and $\mathcal{S}=43$ for $\beta=10$.
			For $\beta=10$, $\mathcal{S}=43$ is the smallest block count for which at least one algorithm achieves $\mathbb{E}_{J}[F]\geq0.95$.
			Columns list the disorder-averaged fidelity $\mathbb{E}_{J}[F]$, mean circuit depth, and single-qubit ($1q$) and two-qubit ($2q$) gate counts.
		}
	\end{table*}
	
	We next examine whether SAP- and OSAP-ma-QAOA remain effective at a larger system size.
	To this end, we evaluate the infidelity trajectories $\mathbb{E}_{J}[1-F]$ for the binary sparse $N=10$ SYK model with $K=16$, as shown in Fig.~\ref{figure5}.
	At $\beta=0.1$ and $1$, the initial ma-QAOA circuit reaches $F\geq99.99\%$ for all disorder realizations at $p=3$.
	At $\beta=10$, the required $p$ varies from $12$ to $17$ across disorder realizations, so we plot the common infidelity trajectories from $\mathcal{S}=384$, corresponding to $p=12$.
	As this figure shows, SAP- and OSAP-ma-QAOA maintain lower infidelities than the two BAP algorithms over the pruning trajectories.
	In addition, Table~\ref{table4} presents $\mathbb{E}_{J}[F]$ and the circuit resources of the four algorithms for a fixed $\mathcal{S}$ at each $\beta$.
	At $\beta=10$, OSAP-ma-QAOA achieves the highest fidelity with circuit resources comparable to those of the other algorithms, and SAP-ma-QAOA also retains high fidelity, whereas the two BAP algorithms show substantially lower fidelities.
	At $\mathcal{S}=43$, OSAP-ma-QAOA removes $88.8\%$--$92.1\%$ of the nonlocal blocks from the initial ma-QAOA circuits with $p=12$--17, while retaining $\mathbb{E}_{J}[F]\simeq0.95$.
	
	\subsection{Effectiveness of small-angle pruning}
	
	We examine the effectiveness of the small-angle selection rule used in SAP- and OSAP-ma-QAOA by comparing them with two random-pruning algorithms.
	First, we consider random sequential pruning, where one block $b_j\in\mathcal{B}_{\rm act}$ is selected randomly and removed at each pruning round, followed by reoptimization over $\bm{\theta}_{\rm act}\setminus\{\theta_j\}$.
	Second, we consider one-shot random pruning, in which a random subset of blocks is removed at once to reach $\mathcal{S}$, followed by reoptimization of the retained parameters.
	These two algorithms are the random counterparts of SAP- and BAP-ma-QAOA, respectively.
	In particular, the comparison between SAP-ma-QAOA and random sequential pruning tests the role of the small-angle selection rule.
	
	\begin{figure*}[t]
		\centering
		\includegraphics[width=\linewidth]{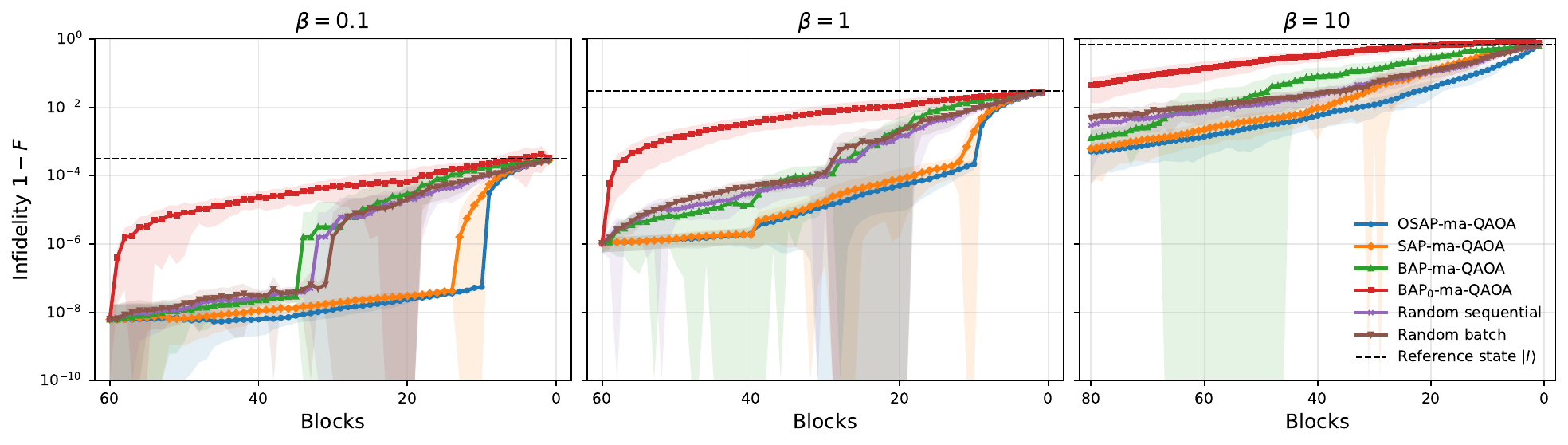}
		\caption{
			\label{figure6}
			Infidelity trajectories of the four pruning algorithms and two random-pruning algorithms for the binary sparse $N=8$ SYK model with $K=10$ at $\beta=0.1$, $1$, and $10$.
			The curves and shaded regions show $\mathbb{E}_{J}[1 - F]$ and $\sigma_{J}[1 - F]$, respectively, averaged over 20 disorder realizations.
			The black dashed lines indicate the disorder-averaged reference infidelity between the exact TFD state and the maximally entangled state $\ket{I}$.
		}
	\end{figure*}
	
	In Fig.~\ref{figure6}, we show $\mathbb{E}_{J}[1-F]$ of SAP-, OSAP-, BAP-, and BAP$_{0}$-ma-QAOA and the two random-pruning algorithms, together with the corresponding $\sigma_{J}[1 - F]$.
	Here, we consider 20 disorder realizations of the binary sparse $N=8$ SYK model with $K=10$.
	This figure shows a consistent performance ordering at all $\beta$ considered: OSAP-ma-QAOA retains the lowest infidelity, followed by SAP-ma-QAOA and the random-pruning algorithms.
	This demonstrates the advantage of the small-angle selection rule.
	As shown by the previous comparison between SAP- and BAP-ma-QAOA in Sec.~\ref{subsec:results}, sequential pruning with intermediate reoptimization provides an additional advantage.
	Taken together, these results indicate that both small-angle selection and sequential pruning are important for maintaining low infidelity.
	
	\section{\label{sec:hardware}Toward hardware-efficient pruning algorithms}
	
	The pruning algorithms considered above involve two practical challenges for implementation on quantum hardware.
	First, the direct-fidelity objective requires explicit preparation of the target TFD state.
	Second, SAP- and OSAP-ma-QAOA require reoptimization after each pruning step, which increases the number of quantum--classical optimization rounds.
	In this section, we address these two aspects by considering an energy-based objective and combined batch--sequential pruning algorithms.
	
	\subsection{\label{energy}Energy-based objective}
	
	VQE provides an alternative objective by optimizing the energy expectation value, which can be directly evaluated through measurements on quantum hardware.
	We therefore examine whether SAP- and OSAP-ma-QAOA retain their pruning performance when the cost function is taken to be $\mathcal{L}=\expval{H_{\rm tot}}$ instead of the infidelity.
	We specifically consider the binary sparse $N=8$ SYK model with $K=10$.
	
	\begin{figure*}[t]
		\centering
		\includegraphics[width=\linewidth]{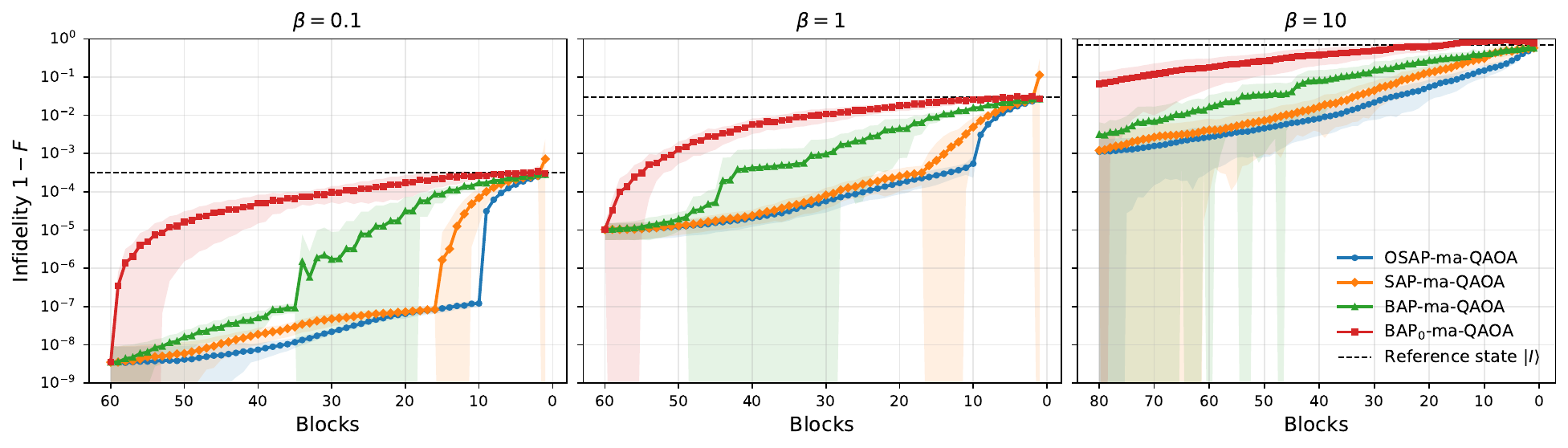}
		\caption{
			\label{figure7}
			Infidelity trajectories of the four pruning algorithms for the binary sparse $N=8$ SYK model with $K=10$ at $\beta=0.1$, $1$, and $10$.
			The variational parameters are optimized using the energy expectation value as the objective function.
			The curves and shaded regions show $\mathbb{E}_{J}[1 - F]$ and $\sigma_{J}[1 - F]$, respectively, averaged over 20 disorder realizations.
			The black dashed lines indicate the disorder-averaged reference infidelity between the exact TFD state and the maximally entangled state $\ket{I}$.
		}
	\end{figure*}
	
	For each disorder realization, we first determine the coupling $g$ in $H_{\rm tot}$ that maximizes the fidelity $\left|\langle \mathrm{TFD}(\beta)|\mathrm{GS}(g)\rangle\right|^{2}$.
	We then optimize the initial ma-QAOA circuit by minimizing $\langle H_{\rm tot}\rangle$ in \eqref{eq:H_tot}, starting from $p=3$ and increasing $p$ when necessary until the resulting state reaches $F\geq99.99\%$.
	The resulting $p$ values are identical to those obtained with the fidelity objective in Sec.~\ref{subsec:results}.
	During pruning, the reoptimization in SAP-, OSAP-, and BAP-ma-QAOA is also performed with this cost function, and the removed block in OSAP-ma-QAOA is selected from $\mathcal{C}$ by the same rule in \eqref{eq:osap_prune_rule}.
	
	The trajectories of $\mathbb{E}_{J}[1-F]$ are shown in Fig.~\ref{figure7} at $\beta=0.1$, $1$, and $10$.
	As this figure shows, SAP- and OSAP-ma-QAOA consistently maintain lower infidelities than the two batch-pruning algorithms across all $\beta$ considered.
	These results show that the advantage of small-angle-based sequential pruning persists when the energy expectation value is used as the objective function and is not restricted to direct fidelity optimization.
	
	\subsection{Combined batch--sequential pruning algorithms}
	
	We next address the repeated reoptimization cost of sequential pruning.
	For a more hardware-efficient implementation, we consider a combined algorithm in which BAP is first applied until the number of retained blocks reaches a prescribed transition block count $\mathcal{S}_{\rm tr}$, after which SAP- or OSAP-ma-QAOA is applied.
	The initial batch pruning reduces the number of reoptimizations.
	After reaching $\mathcal{S}_{\rm tr}$, SAP- or OSAP-ma-QAOA sequentially removes the remaining blocks with reoptimization.
	
	\begin{table*}[t]
		\begin{ruledtabular}
			\begin{tabular}{clccccc}
				$\beta$ & Algorithm & Blocks & $\mathbb{E}_{J}[F]$ & Depth & $1q$ & $2q$ \\
				\hline
				\multirow{6}{*}{$0.1$}
				& SAP        & 1 & 0.999719\,($1\times10^{-6}$) & 16\,(2) & 28\,(3) & 10\,(2) \\
				& OSAP       & 1 & 0.999719\,($1\times10^{-6}$) & 16\,(2) & 27\,(3) & 10\,(3) \\
				& BAP        & 1 & 0.999719\,($1\times10^{-6}$) & 15\,(2) & 27\,(4) & 9\,(2) \\
				& BAP$_{0}$  & 1 & 0.999663\,($1.01\times10^{-4}$) & 15\,(2) & 27\,(4) & 9\,(2) \\
				& BAP + SAP  & 1 & 0.999719\,($1\times10^{-6}$) & 16\,(3) & 28\,(3) & 9\,(3) \\
				& BAP + OSAP & 1 & 0.999719\,($1\times10^{-6}$) & 16\,(2) & 29\,(2) & 9\,(3) \\
				\hline
				\multirow{6}{*}{$1$}
				& SAP        & 1 & 0.972688\,($6.39\times10^{-4}$) & 15\,(2) & 25\,(4) & 9\,(2) \\
				& OSAP       & 1 & 0.972709\,($6.59\times10^{-4}$) & 15\,(2) & 24\,(4) & 9\,(3) \\
				& BAP        & 1 & 0.972596\,($6.67\times10^{-4}$) & 16\,(2) & 28\,(3) & 9\,(2) \\
				& BAP$_{0}$  & 1 & 0.971581\,($1.021\times10^{-3}$) & 16\,(2) & 28\,(3) & 9\,(2) \\
				& BAP + SAP  & 1 & 0.972668\,($6.78\times10^{-4}$) & 16\,(2) & 27\,(4) & 10\,(3) \\
				& BAP + OSAP & 1 & 0.972709\,($6.58\times10^{-4}$) & 15\,(2) & 25\,(4) & 9\,(2) \\
				\hline
				\multirow{6}{*}{$10$}
				& SAP        & 19 & 0.868664\,($7.1575\times10^{-2}$) & 268\,(12) & 293\,(8) & 178\,(14) \\
				& OSAP       & 19 & 0.957391\,($2.0258\times10^{-2}$) & 267\,(14) & 293\,(8) & 175\,(16) \\
				& BAP        & 19 & 0.696494\,($1.13240\times10^{-1}$) & 264\,(17) & 287\,(4) & 173\,(20) \\
				& BAP$_{0}$  & 19 & 0.331946\,($1.80966\times10^{-1}$) & 264\,(17) & 287\,(4) & 173\,(20) \\
				& BAP + SAP  & 19 & 0.860915\,($7.3597\times10^{-2}$) & 267\,(15) & 291\,(9) & 176\,(16) \\
				& BAP + OSAP & 19 & 0.902914\,($6.0646\times10^{-2}$) & 267\,(17) & 292\,(8) & 175\,(19) \\
			\end{tabular}
		\end{ruledtabular}
		\caption{\label{table5}
			Comparison of the six ma-QAOA pruning algorithms for the binary sparse $N=8$ SYK model with $K=10$ at the same target block count $\mathcal{S}$, averaged over 20 disorder realizations.
			For each realization and each algorithm, we evaluate the circuit at fixed $\mathcal{S}$ and report the mean values with the standard deviations in parentheses.
			We set $\mathcal{S}=1$ for $\beta=0.1$ and $1$, and $\mathcal{S}=19$ for $\beta=10$, while the combined algorithms use $\mathcal{S}_{\rm tr}=20$, $20$, and $40$ for $\beta=0.1$, $1$, and $10$, respectively.
			For $\beta=10$, $\mathcal{S}=19$ is the smallest block count for which at least one algorithm achieves $\mathbb{E}_{J}[F]\geq0.95$.
			Columns list the disorder-averaged fidelity $\mathbb{E}_{J}[F]$, mean circuit depth, and single-qubit ($1q$) and two-qubit ($2q$) gate counts.
		}
	\end{table*}
	
	We compare these combined algorithms with SAP-, OSAP-, BAP-, and BAP$_0$-ma-QAOA.
	The comparison at fixed $\mathcal{S}$ for the binary sparse $N=8$ SYK Hamiltonian with $K=10$ is summarized in Table~\ref{table5}.
	For the combined algorithms, we set $\mathcal{S}_{\rm tr}=20$, $20$, and $40$ for $\beta=0.1$, $1$, and $10$, respectively.
	At $\beta=0.1$ and $1$ with $\mathcal{S}=1$, the combined algorithms recover fidelities comparable to those of the sequential algorithms.
	This indicates that the initial batch-pruning stage does not significantly degrade the final result under strong circuit reduction.
	At $\beta=10$ with $\mathcal{S}=19$, however, the combined algorithms achieve higher mean fidelities than BAP- and BAP$_{0}$-ma-QAOA, but lower mean fidelities than their corresponding pure sequential algorithms.
	This shows that sequential reoptimization after the transition is useful, while substantial initial batch pruning can limit the subsequent recovery, especially at low temperature.
	
	\begin{figure}[tb]
		\centering
		\includegraphics[width=\linewidth]{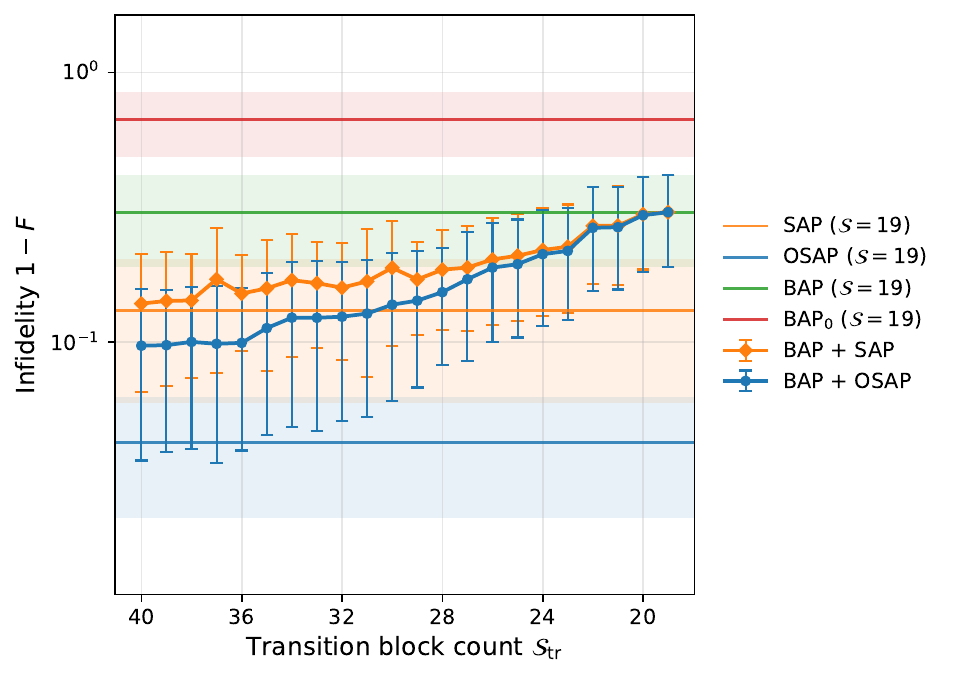}
		\caption{
			\label{figure8}
			Final infidelities at $\mathcal{S}=19$ as a function of $\mathcal{S}_{\rm tr}$ for the binary sparse $N=8$ SYK model with $K=10$ at $\beta=10$.
			The horizontal lines and shaded regions show the reference $\mathbb{E}_{J}[1 - F]$ and $\sigma_{J}[1 - F]$, respectively, for SAP, OSAP, BAP, and BAP$_0$ at $\mathcal{S}=19$, averaged over 20 disorder realizations.
			For BAP+SAP and BAP+OSAP, the curves and vertical error bars indicate $\mathbb{E}_{J}[1 - F]$ and $\sigma_{J}[1 - F]$, respectively, over the same disorder realizations.
		}
	\end{figure}
	
	To examine this tradeoff more directly, Fig.~\ref{figure8} shows $\mathbb{E}_{J}[1-F]$ for BAP+SAP and BAP+OSAP algorithms at fixed $\mathcal{S}=19$ as a function of $\mathcal{S}_{\rm tr}$ at $\beta=10$.
	For comparison, we also show the reference values of $\mathbb{E}_{J}[1-F]$ for SAP-, OSAP-, BAP-, and BAP$_0$-ma-QAOA at $\mathcal{S}=19$ without initial batch pruning.
	The figure shows that BAP+OSAP maintains an infidelity between those of OSAP- and SAP-ma-QAOA when $\mathcal{S}_{\rm tr}$ is not too small.
	As $\mathcal{S}_{\rm tr}$ is reduced further, both combined algorithms approach the BAP reference result.
	This indicates that once too many blocks are removed in the initial batch-pruning step, the subsequent sequential pruning cannot fully restore the lost fidelity.
	Therefore, BAP+OSAP provides a practical compromise between retaining high TFD-state fidelity and reducing the reoptimization cost.
	One may also consider screening the candidates before reoptimizing, which reduces the number of reoptimizations but does not improve the pruning performance over SAP-ma-QAOA, as discussed in Appendix~\ref{sec:ossap}.
	
	Taken together, the energy-based objective considered above and the combined pruning algorithms provide useful ingredients for hardware-oriented TFD-state preparation.
	The energy-based objective replaces direct fidelity evaluation during the variational optimization with measurements of the energy expectation value on quantum hardware.
	In addition, moderate initial batch pruning reduces the number of sequential pruning rounds and thus the cost of quantum--classical reoptimization.
	Therefore, combining the energy-based objective with batch--sequential pruning provides a practical route toward quantum--classical hybrid implementation of high-fidelity TFD-state preparation.

	\section{\label{sec:discussion}Discussion}
	
	We have studied TFD-state preparation and circuit reduction using ma-QAOA for Gaussian and binary SYK models in both dense and sparse cases.
	We find that ma-QAOA achieves higher TFD-state fidelity than standard QAOA, especially at low temperature.
	
	Starting from the optimized initial ma-QAOA circuit, we introduce SAP- and OSAP-ma-QAOA to reduce the number of nonlocal Pauli-string evolutions.
	At each pruning round, SAP-ma-QAOA identifies the nonlocal block with the smallest angle magnitude, removes the corresponding block, and then reoptimizes the remaining variational parameters.
	On the other hand, OSAP-ma-QAOA identifies several candidates with small angle magnitudes, evaluates the post-reoptimization cost for each candidate, and then chooses the block that yields the smallest cost.
	
	We compare the two algorithms with the batch pruning algorithms, BAP- and BAP$_{0}$-ma-QAOA, using the infidelity with the exact TFD state as the cost function.
	We numerically compute the disorder-averaged infidelities for $N=8$ SYK models at three representative temperatures to track the full pruning trajectories for each algorithm.
	The results show that SAP- and OSAP-ma-QAOA retain higher fidelity than batch pruning under strong circuit reduction, particularly at low temperature.
	At fixed target fidelity, sequential pruning also yields more compact circuits.
	
	The comparison between SAP- and BAP-ma-QAOA directly shows that sequential reoptimization at each round is important.
	We also compare our algorithms with random-pruning algorithms, and the comparison shows that identifying blocks with small angle magnitudes is advantageous for retaining the fidelity.
	Taken together, these comparisons indicate that both the small-angle selection rule and sequential pruning with intermediate reoptimization contribute to the performance of SAP-ma-QAOA.
	In addition, the comparison between SAP- and OSAP-ma-QAOA shows that objective-aware pruning further improves the fidelity.
	Thus, the optimized angle magnitude provides an effective criterion for identifying removable blocks, while explicit objective-aware pruning can further improve SAP-ma-QAOA.
	We also confirm that the same qualitative behavior persists for the larger binary sparse $N=10$ SYK model.
	
	For a more hardware-feasible application, we examine our algorithms using the energy expectation value as the cost function, and observe the same qualitative behavior for the binary sparse $N=8$ SYK model.
	We also consider combined algorithms in which batch pruning is first applied and SAP- or OSAP-ma-QAOA is used only after a prescribed transition point.
	These algorithms retain much of the advantage of sequential pruning while reducing the number of reoptimization rounds.
	In particular, the amount of initial batch pruning controls a tradeoff between the reoptimization cost and the final TFD-state fidelity, especially at low temperature.
	Together, the energy-based objective and the combined pruning algorithms provide a route toward quantum--classical hybrid implementation, where each reoptimization requires repeated evaluations of quantum circuits.
	
	Applying SAP- and OSAP-ma-QAOA to other many-body systems would be an interesting direction for future work.
	In addition, implementations on real quantum hardware will be needed to assess their practical performance, particularly for the combined BAP+SAP- and BAP+OSAP-ma-QAOA.
	Beyond the combined algorithms, extending the sequential pruning algorithms toward more hardware-feasible schemes could also be considered.
	For instance, integrating the pruning algorithms with other hybrid quantum--classical approaches could provide a possible route toward practical hardware implementation with reduced quantum--classical optimization cost.
	The reduced TFD-preparation circuits can further be used in finite-temperature quantum simulations, including black-hole-inspired protocols.
	
	\section*{Acknowledgements}
	
	This work was supported by the National Research Foundation of Korea(NRF) grant funded by the Korea government(MSIT)(RS-2026-25477372, RS-2025-02311201, RS-2025-02307394, RS-2024-00445164), and by the Creation of the Quantum Information Science R\&D Ecosystem (Grant No. 2022M3H3A106307411, RS-2023-NR068116) This work was also supported by GIST research fund (Future leading Specialized Resarch Project, 2026 and the ANCHOR program through the Gwangju ANCHOR Center, funded by the Ministry of Education(MOE) and the Jeonnam-Gwangju Special Metropolitan City, Republic of Korea.(2026-ANCHOR-05-001)

	\appendix
	
	\section*{Appendices}
	
	\section{\label{sec:single_realization}Circuit resources for a single disorder realization}
	
	\begin{table*}[t]
		\centering
		\footnotesize
		\begin{ruledtabular}
			\begin{tabular}{llccccc}
				Model & $\beta$ & Blocks
				& SAP $(F,D,2q)$
				& OSAP $(F,D,2q)$
				& BAP $(F,D,2q)$
				& BAP$_0$ $(F,D,2q)$ \\
				\hline
				\multirow{3}{*}{Gaussian dense}
				& $0.1$ & 1  & $(0.999774,17,12)$  & $(0.999774,14,12)$  & $(0.999750,16,12)$  & $(0.999194,16,12)$ \\
				& $1$   & 1  & $(0.976565,13,6)$   & $(0.977732,14,12)$  & $(0.977732,17,12)$  & $(0.975150,17,12)$ \\
				& $10$  & 27 & $(0.900126,390,264)$ & $(0.903129,350,260)$ & $(0.669229,382,258)$ & $(0.449123,382,258)$ \\
				\hline
				\multirow{3}{*}{Gaussian sparse}
				& $0.1$ & 1  & $(0.999717,16,10)$  & $(0.999759,10,6)$   & $(0.999759,13,6)$   & $(0.999753,13,6)$ \\
				& $1$   & 1  & $(0.973337,15,10)$  & $(0.976845,10,6)$   & $(0.976845,13,6)$   & $(0.975685,13,6)$ \\
				& $10$  & 24 & $(0.833507,332,220)$ & $(0.907946,300,216)$ & $(0.738726,331,220)$ & $(0.447241,331,220)$ \\
				\hline
				\multirow{3}{*}{Binary dense}
				& $0.1$ & 1  & $(0.999693,12,6)$   & $(0.999693,13,6)$   & $(0.999689,13,6)$   & $(0.999269,13,6)$ \\
				& $1$   & 1  & $(0.971194,13,6)$   & $(0.971365,12,6)$   & $(0.971355,15,10)$  & $(0.970519,15,10)$ \\
				& $10$  & 48 & $(0.845357,669,448)$ & $(0.901974,661,442)$ & $(0.795299,663,444)$ & $(0.442123,663,444)$ \\
				\hline
				\multirow{3}{*}{Binary sparse}
				& $0.1$ & 1  & $(0.999719,12,6)$   & $(0.999719,10,8)$   & $(0.999719,15,8)$   & $(0.999713,15,8)$ \\
				& $1$   & 1  & $(0.972901,15,8)$   & $(0.972901,12,8)$   & $(0.972895,15,8)$   & $(0.972085,15,8)$ \\
				& $10$  & 25 & $(0.729645,295,168)$ & $(0.959548,294,202)$ & $(0.676972,306,184)$ & $(0.233507,306,184)$ \\
			\end{tabular}
		\end{ruledtabular}
		\caption{\label{table6}
			Comparison of the four ma-QAOA pruning algorithms for the four $N=8$ SYK models at the same target block count $\mathcal{S}$ for a single disorder realization of each model.
			For the sparse models, Gaussian sparse uses $K=32$ and binary sparse uses $K=10$.
			For each model and each $\beta$, we evaluate the four algorithms at fixed $\mathcal{S}$ and report $(F,D,2q)$, where $D$ denotes the circuit depth.
			We set $\mathcal{S}=1$ for $\beta=0.1$ and $1$, and $\mathcal{S}=27$, $24$, $48$, and $25$ for the Gaussian dense, Gaussian sparse, binary dense, and binary sparse models at $\beta=10$, respectively.
			For $\beta=10$, $\mathcal{S}$ is the smallest block count for which at least one algorithm achieves $F\geq0.9$, except for the binary sparse SYK model, for which $F\geq0.95$ is used.
		}
	\end{table*}
	
	Table~\ref{table6} summarizes the fidelities and circuit resources of SAP-, OSAP-, BAP-, and BAP$_{0}$-ma-QAOA for a single disorder realization in each SYK model.
	Here we report $(F,D,2q)$, corresponding to the fidelity, circuit depth, and number of two-qubit gates, respectively, at fixed $\mathcal{S}$ for each $\beta$.
	At $\beta=0.1$ and $1$, the differences among the four algorithms remain small even when the circuit is reduced to a single block.
	In contrast, a clear separation between sequential and batch pruning appears at $\beta=10$, where the performance differences persist under strong pruning.
	While the relative performance varies across disorder realizations, OSAP-ma-QAOA in the present case achieves the highest fidelity for all four SYK models.
	These results indicate that sequential reoptimization is important for retaining the TFD-state fidelity under strong pruning at low temperature, while a systematic comparison between SAP- and OSAP-ma-QAOA requires averaging over disorder realizations, as shown in Figs.~\ref{figure3} and~\ref{figure4}.
	
	\section{\label{sec:pool_size}Dependence on the candidate pool size in OSAP-ma-QAOA}
	
	In OSAP-ma-QAOA, the candidate pool size $N_{\mathcal C}$ determines the number of candidate blocks whose post-reoptimization costs are evaluated at each pruning round.
	Increasing $N_{\mathcal C}$ allows more candidate blocks to be considered for removal, while increasing the number of reoptimizations required at each round.
	
\begin{figure}[t]
		\centering
		\includegraphics[width=1.0\linewidth]{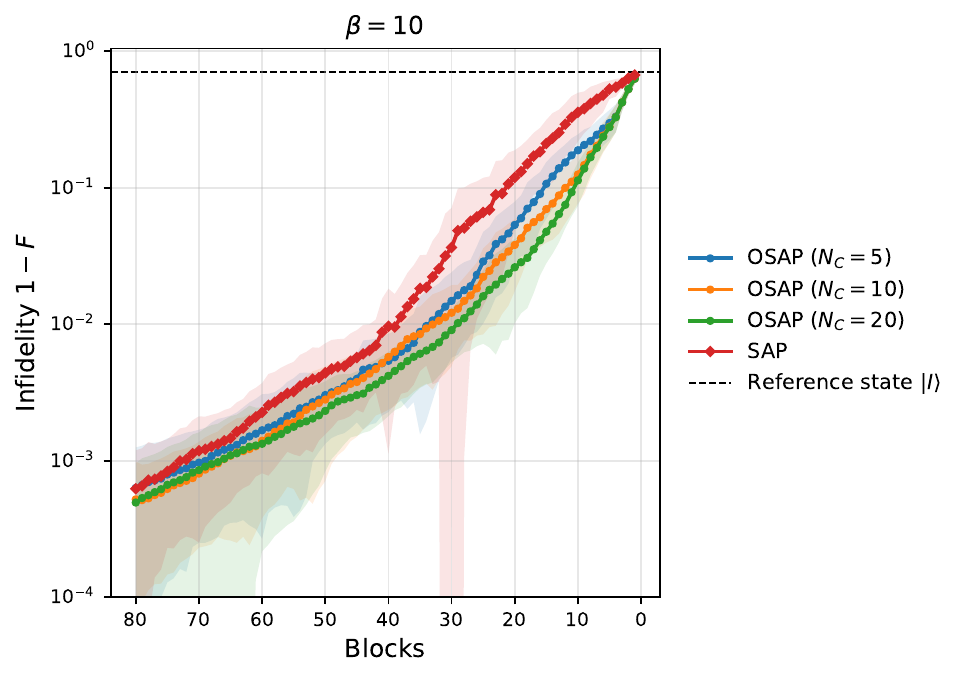}
		\caption{\label{figure9}
			Trajectories of $\mathbb{E}_{J}[1-F]$ (solid) and $\sigma_{J}[1 - F]$ (shaded) for OSAP-ma-QAOA, averaged over 20 disorder realizations of the binary sparse $N=8$ SYK model with $K=10$ at $\beta=10$.
			For OSAP-ma-QAOA, we consider $N_{\mathcal C}=5$, $10$, and $20$, together with $N_{\mathcal C}=1$, which effectively corresponds to SAP-ma-QAOA.
			The black dashed line indicates the disorder-averaged reference infidelity between the exact TFD state and the maximally entangled state $\ket{I}$.
		}
	\end{figure}

	We examine the dependence on $N_{\mathcal C}$ for the binary sparse $N=8$ SYK model with $K=10$ at $\beta=10$, where the difference between SAP- and OSAP-ma-QAOA is most pronounced.
	We compare OSAP-ma-QAOA with $N_{\mathcal C}=5$, $10$, and $20$, together with SAP-ma-QAOA, which effectively corresponds to $N_{\mathcal C}=1$.
	Figure~\ref{figure9} shows the resulting trajectories of $\mathbb{E}_{J}[1-F]$ as a function of $\mathcal{S}$.
	For relatively large $\mathcal{S}$, the three choices of $N_{\mathcal C}$ give similar infidelities.
	As $\mathcal{S}$ decreases, a larger $N_{\mathcal C}$ generally gives lower infidelity until hard pruning.
	Thus, a larger $N_{\mathcal C}$ further improves the performance of OSAP-ma-QAOA under strong circuit reduction.
	On the other hand, a larger $N_{\mathcal C}$ increases the reoptimization cost.
	Thus, we use $N_{\mathcal C}=10$ throughout the main text as a practical compromise between the performance of OSAP-ma-QAOA and the computational cost.
	
	\section{Objective-screened sequential angle pruning}
	\label{sec:ossap}
	
	\begin{figure*}[tb]
		\centering
		\includegraphics[width=\linewidth]{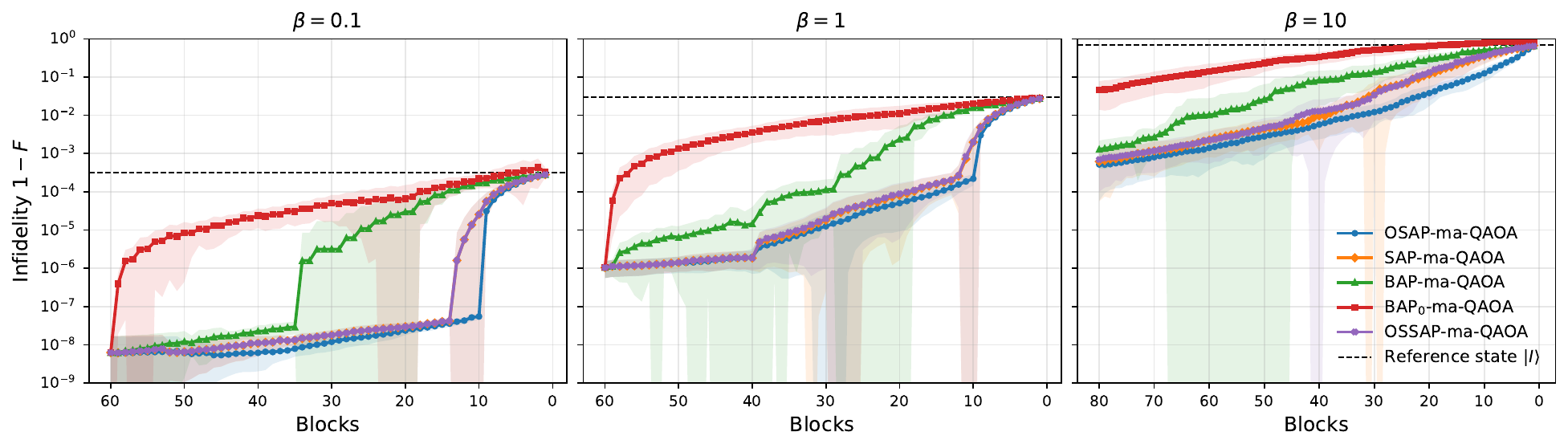}
		\caption{
			Infidelity trajectories of the five pruning algorithms for the binary sparse $N=8$ SYK model with $K=10$ at $\beta=0.1$, $1$, and $10$.
			The curves and shaded regions show $\mathbb{E}_{J}[1 - F]$ and $\sigma_{J}[1 - F]$, respectively, averaged over 20 disorder realizations.
			The black dashed lines indicate the disorder-averaged reference infidelity between the exact TFD state and the maximally entangled state $\ket{I}$.
		}
		\label{fig:ossap_comparison}
	\end{figure*}
	
	Although OSAP-ma-QAOA improves the pruning performance by using the post-reoptimization cost for selecting the removed block, it requires reoptimization for every candidate block at each pruning round, so the reoptimization cost increases with the candidate-pool size $N_{\mathcal C}$.
	To reduce this cost, we consider objective-screened SAP-ma-QAOA (OSSAP-ma-QAOA), which extends SAP-ma-QAOA by screening small-angle candidates using their pre-reoptimization costs.
	Unlike OSAP-ma-QAOA, OSSAP-ma-QAOA performs only one reoptimization after the removed block is selected.
	
	Starting from the optimized initial ma-QAOA ansatz, OSSAP-ma-QAOA forms the same small-angle candidate pool $\mathcal{C}$ as OSAP-ma-QAOA at each pruning round.
	For each candidate block $b_j\in\mathcal{C}$, we tentatively remove $b_j$ while keeping all remaining parameters fixed at their current optimized values.
	We then evaluate the corresponding pre-reoptimization cost,
	\begin{equation}
		\label{eq:ossap_cost}
		\mathcal{L}_{j}^{\rm pre}
		=
		\left.\mathcal{L}(\bm{\theta}_{\rm act})\right|_{\theta_j=0},
	\end{equation}
	without reoptimizing the remaining parameters.
	The candidate to be removed is selected as
	\begin{equation}
		\label{eq:ossap_prune_rule}
		j^{\star} = \argmin_{j:\,b_j\in\mathcal{C}}\mathcal{L}_{j}^{\rm pre}.
	\end{equation}
	After removing $b_{j^\star}$, we update $\mathcal{B}_{\rm act}\leftarrow\mathcal{B}_{\rm act}\setminus\{b_{j^\star}\}$ and $\bm{\theta}_{\rm act}\leftarrow\bm{\theta}_{\rm act}\setminus\{\theta_{j^\star}\}$, and reoptimize $\bm{\theta}_{\rm act}$ once.
	The resulting angles are then used to form the next $\mathcal{C}$.
	The procedure is repeated until $\mathcal{N}(\mathcal{B}_{\rm act})=\mathcal{S}$.
	
	The difference between OSSAP- and OSAP-ma-QAOA lies in how the reoptimization is used for candidate selection.
	For a given $\mathcal{C}$, OSAP-ma-QAOA performs $\mathcal{N}(\mathcal{C})$ reoptimizations at each pruning round and selects the removed block according to the post-reoptimization costs.
	In contrast, OSSAP-ma-QAOA evaluates the $\mathcal{N}(\mathcal{C})$ pre-reoptimization costs and performs only one reoptimization after the removed block is selected.
	Thus, OSSAP-ma-QAOA reduces the number of reoptimizations required for objective-aware pruning.
	
	In Fig.~\ref{fig:ossap_comparison}, we show $\mathbb{E}_{J}[1-F]$ for SAP-, OSAP-, BAP-, BAP$_{0}$-, and OSSAP-ma-QAOA, together with the corresponding $\sigma_{J}[1 - F]$.
	Here, we consider the binary sparse $N=8$ SYK model with $K=10$ at $\beta=0.1$, $1$, and $10$, using 20 disorder realizations.
	For all three temperatures, OSSAP-ma-QAOA closely follows the trajectories of SAP-ma-QAOA.
	
	We also find that OSSAP-ma-QAOA selects the nonlocal block with the smallest angle magnitude in most pruning rounds, while OSAP-ma-QAOA more frequently selects a different candidate.
	This behavior is consistent with the local cost analysis discussed in Sec.~\ref{sec:pruning}.
	Before reoptimization at each pruning round, the cost change caused by removing a block is approximately given by $\Delta\mathcal{L}_{j}^{\rm pre}\sim\mathcal{H}_{jj}|\theta_j^{\ast}|^2$.
	Therefore, when the variation of $\mathcal{H}_{jj}$ among the small-angle candidates is not large, minimizing the pre-reoptimization cost in OSSAP-ma-QAOA is expected to select the same block as the small-angle criterion in SAP-ma-QAOA.
	
	To compare the computational costs of the sequential pruning algorithms, let $C_{\rm eval}$ denote the cost of a single objective-function evaluation and $C_{\rm opt}$ the cost of one reoptimization of the remaining parameters.
	For a candidate pool of size $\mathcal{N}(\mathcal{C})$, SAP-ma-QAOA requires one reoptimization at each pruning round, while OSSAP-ma-QAOA requires $\mathcal{N}(\mathcal{C})$ objective-function evaluations followed by one reoptimization.
	In contrast, OSAP-ma-QAOA requires one reoptimization for each candidate block.
	Thus, the computational costs per pruning round can be expressed schematically as
	\begin{equation}
		\begin{aligned}
			C_{\rm SAP} &\sim C_{\rm opt},\\
			C_{\rm OSSAP} &\sim \mathcal{N}(\mathcal{C})C_{\rm eval}+C_{\rm opt},\\
			C_{\rm OSAP} &\sim \mathcal{N}(\mathcal{C})C_{\rm opt}.
		\end{aligned}
		\label{eq:pruning_cost}
	\end{equation}
	Since a reoptimization generally requires multiple objective-function evaluations, $C_{\rm opt}$ is typically larger than $C_{\rm eval}$.
	OSSAP-ma-QAOA therefore reduces the $\mathcal{N}(\mathcal{C})$ reoptimizations required by OSAP-ma-QAOA to a single reoptimization, at the cost of $\mathcal{N}(\mathcal{C})$ direct objective-function evaluations.
	However, because OSSAP- and SAP-ma-QAOA show similar performance while $C_{\rm SAP} < C_{\rm OSSAP}$, our results indicate that evaluating the objective function before reoptimization does not improve the pruning performance over SAP-ma-QAOA.

	\bibliographystyle{jhep} % use this if jhep.bst is available locally
	\bibliography{ref}

@article{Caceres2021,
	author = {C{\'a}ceres, Elena and Misobuchi, Anderson and Pimentel, Rafael},
	date = {2021/11/04},
	doi = {10.1007/JHEP11(2021)015},
	id = {C{\'a}ceres2021},
	isbn = {1029-8479},
	journal = {Journal of High Energy Physics},
	number = {11},
	pages = {15},
	title = {Sparse SYK and traversable wormholes},
	url = {https://doi.org/10.1007/JHEP11(2021)015},
	volume = {2021},
	year = {2021}
}

@article{PhysRevA.104.012427,
	title = {Variational preparation of the thermofield double state of the Sachdev-Ye-Kitaev model},
	author = {Su, Vincent Paul},
	journal = {Phys. Rev. A},
	volume = {104},
	issue = {1},
	pages = {012427},
	numpages = {13},
	year = {2021},
	month = {Jul},
	publisher = {American Physical Society},
	doi = {10.1103/PhysRevA.104.012427},
	url = {https://link.aps.org/doi/10.1103/PhysRevA.104.012427}
}

@article{Jafferis2022,
	author = {Jafferis, Daniel and Zlokapa, Alexander and Lykken, Joseph D. and Kolchmeyer, David K. and Davis, Samantha I. and Lauk, Nikolai and Neven, Hartmut and Spiropulu, Maria},
	date = {2022/12/01},
	doi = {10.1038/s41586-022-05424-3},
	id = {Jafferis2022},
	isbn = {1476-4687},
	journal = {Nature},
	number = {7938},
	pages = {51--55},
	title = {Traversable wormhole dynamics on a quantum processor},
	url = {https://doi.org/10.1038/s41586-022-05424-3},
	volume = {612},
	year = {2022}}

@article{PhysRevLett.70.3339,
	title = {Gapless spin-fluid ground state in a random quantum Heisenberg magnet},
	author = {Sachdev, Subir and Ye, Jinwu},
	journal = {Phys. Rev. Lett.},
	volume = {70},
	issue = {21},
	pages = {3339--3342},
	numpages = {0},
	year = {1993},
	month = {May},
	publisher = {American Physical Society},
	doi = {10.1103/PhysRevLett.70.3339},
	url = {https://link.aps.org/doi/10.1103/PhysRevLett.70.3339}
}

@article{Maldacena2016,
	author = {Maldacena, Juan and Shenker, Stephen H. and Stanford, Douglas},
	date = {2016/08/17},
	doi = {10.1007/JHEP08(2016)106},
	id = {Maldacena2016},
	isbn = {1029-8479},
	journal = {Journal of High Energy Physics},
	number = {8},
	pages = {106},
	title = {A bound on chaos},
	url = {https://doi.org/10.1007/JHEP08(2016)106},
	volume = {2016},
	year = {2016}}

@article{PhysRevLett.117.111601,
	title = {Chaos in ${\mathrm{AdS}}_{2}$ Holography},
	author = {Jensen, Kristan},
	journal = {Phys. Rev. Lett.},
	volume = {117},
	issue = {11},
	pages = {111601},
	numpages = {6},
	year = {2016},
	month = {Sep},
	publisher = {American Physical Society},
	doi = {10.1103/PhysRevLett.117.111601},
	url = {https://link.aps.org/doi/10.1103/PhysRevLett.117.111601}
}

@article{Maldacena:2018lmt,
	author = "Maldacena, Juan and Qi, Xiao-Liang",
	title = "{Eternal traversable wormhole}",
	eprint = "1804.00491",
	archivePrefix = "arXiv",
	primaryClass = "hep-th",
	month = "4",
	year = "2018"
}

@misc{xu2020sparsemodelquantumholography,
	title={A Sparse Model of Quantum Holography}, 
	author={Shenglong Xu and Leonard Susskind and Yuan Su and Brian Swingle},
	year={2020},
	eprint={2008.02303},
	archivePrefix={arXiv},
	primaryClass={cond-mat.str-el},
	url={https://arxiv.org/abs/2008.02303}, 
}

@article{PhysRevD.103.106002,
	title = {Sparse Sachdev-Ye-Kitaev model, quantum chaos, and gravity duals},
	author = {Garc\'{\i}a-Garc\'{\i}a, Antonio M. and Jia, Yiyang and Rosa, Dario and Verbaarschot, Jacobus J. M.},
	journal = {Phys. Rev. D},
	volume = {103},
	issue = {10},
	pages = {106002},
	numpages = {28},
	year = {2021},
	month = {May},
	publisher = {American Physical Society},
	doi = {10.1103/PhysRevD.103.106002},
	url = {https://link.aps.org/doi/10.1103/PhysRevD.103.106002}
}

@article{PhysRevB.107.L081103,
	title = {Binary-coupling sparse Sachdev-Ye-Kitaev model: An improved model of quantum chaos and holography},
	author = {Tezuka, Masaki and Oktay, Onur and Rinaldi, Enrico and Hanada, Masanori and Nori, Franco},
	journal = {Phys. Rev. B},
	volume = {107},
	issue = {8},
	pages = {L081103},
	numpages = {7},
	year = {2023},
	month = {Feb},
	publisher = {American Physical Society},
	doi = {10.1103/PhysRevB.107.L081103},
	url = {https://link.aps.org/doi/10.1103/PhysRevB.107.L081103}
}

@article{Orman2025,
	author = {Orman, Patrick and Gharibyan, Hrant and Preskill, John},
	date = {2025/02/26},
	doi = {10.1007/JHEP02(2025)173},
	id = {Orman2025},
	isbn = {1029-8479},
	journal = {Journal of High Energy Physics},
	number = {2},
	pages = {173},
	title = {Quantum chaos in the sparse SYK model},
	url = {https://doi.org/10.1007/JHEP02(2025)173},
	volume = {2025},
	year = {2025}}

@article{Kandala2017,
	author = {Kandala, Abhinav and Mezzacapo, Antonio and Temme, Kristan and Takita, Maika and Brink, Markus and Chow, Jerry M. and Gambetta, Jay M.},
	date = {2017/09/01},
	doi = {10.1038/nature23879},
	id = {Kandala2017},
	isbn = {1476-4687},
	journal = {Nature},
	number = {7671},
	pages = {242--246},
	title = {Hardware-efficient variational quantum eigensolver for small molecules and quantum magnets},
	url = {https://doi.org/10.1038/nature23879},
	volume = {549},
	year = {2017}}

@article{PhysRevD.94.106002,
	title = {Remarks on the Sachdev-Ye-Kitaev model},
	author = {Maldacena, Juan and Stanford, Douglas},
	journal = {Phys. Rev. D},
	volume = {94},
	issue = {10},
	pages = {106002},
	numpages = {43},
	year = {2016},
	month = {Nov},
	publisher = {American Physical Society},
	doi = {10.1103/PhysRevD.94.106002},
	url = {https://link.aps.org/doi/10.1103/PhysRevD.94.106002}
}

@misc{Kitaev2015,
	author       = {Kitaev, Alexei},
	title        = {A Simple Model of Quantum Holography},
	howpublished = {KITP Strings Seminar and Entanglement 2015 Program},
	year         = {2015},
	note         = {Talks given on February 12, April 7, and May 27, 2015},
	url          = {http://online.kitp.ucsb.edu/online/entangled15/}
}

@article{Granet2026,
	author = {Granet, Etienne and Kikuchi, Yuta and Dreyer, Henrik and Rinaldi, Enrico},
	date = {2026/02/23},
	doi = {10.1038/s41534-026-01206-1},
	id = {Granet2026},
	isbn = {2056-6387},
	journal = {npj Quantum Information},
	number = {1},
	pages = {43},
	title = {Simulating sparse SYK model with a randomized algorithm on a trapped-ion quantum computer},
	url = {https://doi.org/10.1038/s41534-026-01206-1},
	volume = {12},
	year = {2026}}

@article{PhysRevLett.126.030602,
	title = {Many-Body Chaos in the Sachdev-Ye-Kitaev Model},
	author = {Kobrin, Bryce and Yang, Zhenbin and Kahanamoku-Meyer, Gregory D. and Olund, Christopher T. and Moore, Joel E. and Stanford, Douglas and Yao, Norman Y.},
	journal = {Phys. Rev. Lett.},
	volume = {126},
	issue = {3},
	pages = {030602},
	numpages = {6},
	year = {2021},
	month = {Jan},
	publisher = {American Physical Society},
	doi = {10.1103/PhysRevLett.126.030602},
	url = {https://link.aps.org/doi/10.1103/PhysRevLett.126.030602}
}

@article{Kundu_2025,
	doi = {10.1088/2632-2153/ade361},
	url = {https://doi.org/10.1088/2632-2153/ade361},
	year = {2025},
	month = {jun},
	publisher = {IOP Publishing},
	volume = {6},
	number = {2},
	pages = {025066},
	author = {Kundu, Akash},
	title = {Improving thermal state preparation of Sachdev–Ye–Kitaev model with reinforcement learning on quantum hardware},
	journal = {Machine Learning: Science and Technology}
}

@article{PhysRevLett.123.220502,
	title = {Variational Thermal Quantum Simulation via Thermofield Double States},
	author = {Wu, Jingxiang and Hsieh, Timothy H.},
	journal = {Phys. Rev. Lett.},
	volume = {123},
	issue = {22},
	pages = {220502},
	numpages = {6},
	year = {2019},
	month = {Nov},
	publisher = {American Physical Society},
	doi = {10.1103/PhysRevLett.123.220502},
	url = {https://link.aps.org/doi/10.1103/PhysRevLett.123.220502}
}

@article{10.1093/ptep/ptw124,
	author = {Maldacena, Juan and Stanford, Douglas and Yang, Zhenbin},
	title = {Conformal symmetry and its breaking in two-dimensional nearly anti-de Sitter space},
	journal = {Progress of Theoretical and Experimental Physics},
	volume = {2016},
	number = {12},
	pages = {12C104},
	year = {2016},
	month = {12},
	issn = {2050-3911},
	doi = {10.1093/ptep/ptw124},
	url = {https://doi.org/10.1093/ptep/ptw124},
}

@article{Preskill2018quantumcomputingin,
	doi = {10.22331/q-2018-08-06-79},
	url = {https://doi.org/10.22331/q-2018-08-06-79},
	title = {Quantum {C}omputing in the {NISQ} era and beyond},
	author = {Preskill, John},
	journal = {{Quantum}},
	issn = {2521-327X},
	publisher = {{Verein zur F{\"{o}}rderung des Open Access Publizierens in den Quantenwissenschaften}},
	volume = {2},
	pages = {79},
	month = aug,
	year = {2018}
}

@misc{Farhi2014qaoa,
	author = {Edward Farhi and Jeffrey Goldstone and Sam Gutmann},
	title = {A Quantum Approximate Optimization Algorithm},
	year = {2014},
	eprint = {1411.4028},
	archivePrefix = {arXiv},
	primaryClass = {quant-ph},
}

@article{Hadfield2019,
	author = {Hadfield, Stuart and Wang, Zhihui and O'Gorman, Bryan and Rieffel, Eleanor G. and Venturelli, Davide and Biswas, Rupak},
	title = {From the Quantum Approximate Optimization Algorithm to a Quantum Alternating Operator Ansatz},
	journal = {Algorithms},
	volume = {12},
	number = {2},
	pages = {34},
	year = {2019},
	doi = {10.3390/a12020034},
}

@article{Herrman2022,
	author = {Herrman, Rebekah and Lotshaw, Phillip C. and Ostrowski, James and Humble, Travis S. and Siopsis, George},
	title = {Multi-angle quantum approximate optimization algorithm},
	journal = {Scientific Reports},
	volume = {12},
	pages = {6781},
	year = {2022},
	doi = {10.1038/s41598-022-10555-8},
}

@article{Bang:2026eof,
	author = "Bang, Jeongho and Byun, Moongul and Cho, Kyoungho and Kim, Keun-Young and Lee, Hyeonsoo",
	title = "{Hayden--Preskill recovery at finite temperature on a quantum processor: dynamics and initial state from the SYK model}",
	eprint = "2607.28486",
	archivePrefix = "arXiv",
	primaryClass = "hep-th",
	month = "7",
	year = "2026"
}

@article{Byun:2026ewk,
	author = "Byun, Moongul and Kim, Keun-Young and Lee, Hyeonsoo",
	title = "{Quantum simulation of traversable-wormhole-inspired quantum teleportation in a chaotic binary sparse SYK model}",
	eprint = "2604.10090",
	archivePrefix = "arXiv",
	primaryClass = "hep-th",
	month = "4",
	year = "2026"
}

@article{Sagastizabal2021,
	author = {Sagastizabal, R. and Premaratne, S. P. and Klaver, B. A. and Rol, M. A. and Neg{\^\i}rneac, V. and Moreira, M. S. and Zou, X. and Johri, S. and Muthusubramanian, N. and Beekman, M. and Zachariadis, C. and Ostroukh, V. P. and Haider, N. and Bruno, A. and Matsuura, A. Y. and DiCarlo, L.},
	date = {2021/08/20},
	doi = {10.1038/s41534-021-00468-1},
	id = {Sagastizabal2021},
	isbn = {2056-6387},
	journal = {npj Quantum Information},
	number = {1},
	pages = {130},
	title = {Variational preparation of finite-temperature states on a quantum computer},
	url = {https://doi.org/10.1038/s41534-021-00468-1},
	volume = {7},
	year = {2021}}

@article{PhysRevA.111.012432,
	title = {Hamiltonian forging of a thermofield double},
	author = {Fa\'{\i}lde, Daniel and Santos-Su\'arez, Juan and Herrera-Mart\'{\i}, David A. and Mas, Javier},
	journal = {Phys. Rev. A},
	volume = {111},
	issue = {1},
	pages = {012432},
	numpages = {13},
	year = {2025},
	month = {Jan},
	publisher = {American Physical Society},
	doi = {10.1103/PhysRevA.111.012432},
	url = {https://link.aps.org/doi/10.1103/PhysRevA.111.012432}
}

@article{PhysRevResearch.2.023074,
	title = {Quantum optimization with a novel Gibbs objective function and ansatz architecture search},
	author = {Li, Li and Fan, Minjie and Coram, Marc and Riley, Patrick and Leichenauer, Stefan},
	journal = {Phys. Rev. Res.},
	volume = {2},
	issue = {2},
	pages = {023074},
	numpages = {10},
	year = {2020},
	month = {Apr},
	publisher = {American Physical Society},
	doi = {10.1103/PhysRevResearch.2.023074},
	url = {https://link.aps.org/doi/10.1103/PhysRevResearch.2.023074}
}

@article{Sim_2021,
	doi = {10.1088/2058-9565/abe107},
	url = {https://doi.org/10.1088/2058-9565/abe107},
	year = {2021},
	month = {mar},
	publisher = {IOP Publishing},
	volume = {6},
	number = {2},
	pages = {025019},
	author = {Sim, Sukin and Romero, Jonathan and Gonthier, Jérôme F and Kunitsa, Alexander A},
	title = {Adaptive pruning-based optimization of parameterized quantum circuits},
	journal = {Quantum Science and Technology}
}

@INPROCEEDINGS{Shi9996634,
	author={Shi, Kaiyan and Herrman, Rebekah and Shaydulin, Ruslan and Chakrabarti, Shouvanik and Pistoia, Marco and Larson, Jeffrey},
	booktitle={2022 IEEE/ACM 7th Symposium on Edge Computing (SEC)}, 
	title={Multiangle QAOA Does Not Always Need All Its Angles}, 
	year={2022},
	volume={},
	number={},
	pages={414-419},
	doi={10.1109/SEC54971.2022.00062}}

@article{10.1021/acs.jctc.5c00535,
	author = {Vaquero-Sabater, Nonia and Carreras, Abel and Casanova, David},
	title = {Pruned-ADAPT-VQE: Compacting Molecular Ans{\"a}tze by Removing Irrelevant Operators},
	journal = {Journal of Chemical Theory and Computation},
	volume = {21},
	number = {18},
	pages = {8720-8728},
	year = {2025},
	month = {09},
	issn = {1549-9618},
	doi = {10.1021/acs.jctc.5c00535},
	url = {https://doi.org/10.1021/acs.jctc.5c00535},
}

@inproceedings{Kim:2026pnn,
	author = "Kim, Hyunwoo and Lee, Youngseok",
	title = "{SAFE ma-QAOA: Surrogate-Assisted and Fine-Tuning Enhanced Multi-Angle QAOA with Parameter Distillation}",
	booktitle = "{2026 IEEE International Conference on Quantum Computing and Engineering}",
	eprint = "2605.23377",
	archivePrefix = "arXiv",
	primaryClass = "quant-ph",
	month = "5",
	year = "2026"
}

@article{Warren:2022evv,
	author = "Warren, Ada and Zhu, Linghua and Mayhall, Nicholas J. and Barnes, Edwin and Economou, Sophia E.",
	title = "{Adaptive variational algorithms for quantum Gibbs state preparation}",
	eprint = "2203.12757",
	archivePrefix = "arXiv",
	primaryClass = "quant-ph",
	month = "3",
	year = "2022"
}

@INPROCEEDINGS{Premaratne:9259931,
	author={Premaratne, Shavindra P. and Matsuura, A. Y.},
	booktitle={2020 IEEE International Conference on Quantum Computing and Engineering (QCE)}, 
	title={Engineering a Cost Function for Real-world Implementation of a Variational Quantum Algorithm}, 
	year={2020},
	volume={},
	number={},
	pages={278-285},
	doi={10.1109/QCE49297.2020.00042}}

@article{Peruzzo2014,
	author = {Peruzzo, Alberto and McClean, Jarrod and Shadbolt, Peter and Yung, Man-Hong and Zhou, Xiao-Qi and Love, Peter J. and Aspuru-Guzik, Al{\'a}n and O'Brien, Jeremy L.},
	date = {2014/07/23},
	doi = {10.1038/ncomms5213},
	id = {Peruzzo2014},
	isbn = {2041-1723},
	journal = {Nature Communications},
	number = {1},
	pages = {4213},
	title = {A variational eigenvalue solver on a photonic quantum processor},
	url = {https://doi.org/10.1038/ncomms5213},
	volume = {5},
	year = {2014}}

@article{10.1007/s11128-021-03001-7,
	author = {Herrman, Rebekah and Ostrowski, James and Humble, Travis S. and Siopsis, George},
	title = {Lower bounds on circuit depth of the quantum approximate optimization algorithm},
	year = {2021},
	issue_date = {Feb 2021},
	publisher = {Kluwer Academic Publishers},
	address = {USA},
	volume = {20},
	number = {2},
	issn = {1570-0755},
	url = {https://doi.org/10.1007/s11128-021-03001-7},
	doi = {10.1007/s11128-021-03001-7},
	journal = {Quantum Information Processing},
	month = feb,
	numpages = {17}
}

@article{PhysRevApplied.16.054035,
	title = {Variational Quantum Gibbs State Preparation with a Truncated Taylor Series},
	author = {Wang, Youle and Li, Guangxi and Wang, Xin},
	journal = {Phys. Rev. Appl.},
	volume = {16},
	issue = {5},
	pages = {054035},
	numpages = {17},
	year = {2021},
	month = {Nov},
	publisher = {American Physical Society},
	doi = {10.1103/PhysRevApplied.16.054035},
	url = {https://link.aps.org/doi/10.1103/PhysRevApplied.16.054035}
}

@article{
	Zhu:2020,
	author = {D. Zhu  and S. Johri  and N. M. Linke  and K. A. Landsman  and C. Huerta Alderete  and N. H. Nguyen  and A. Y. Matsuura  and T. H. Hsieh  and C. Monroe },
	title = {Generation of thermofield double states and critical ground states with a quantum computer},
	journal = {Proceedings of the National Academy of Sciences},
	volume = {117},
	number = {41},
	pages = {25402-25406},
	year = {2020},
	doi = {10.1073/pnas.2006337117},
	URL = {https://www.pnas.org/doi/abs/10.1073/pnas.2006337117}}

@article{Guo_2023,
	doi = {10.1088/1674-1056/aca7f3},
	url = {https://doi.org/10.1088/1674-1056/aca7f3},
	year = {2023},
	month = {jan},
	publisher = {Chinese Physical Society and IOP Publishing Ltd},
	volume = {32},
	number = {1},
	pages = {010307},
	author = {Guo, Xue-Yi and Li, Shang-Shu and Xiao, Xiao and Xiang, Zhong-Cheng and Ge, Zi-Yong and Li, He-Kang and Song, Peng-Tao and Peng, Yi and Wang, Zhan and Xu, Kai and Zhang, Pan and Wang, Lei and Zheng, Dong-Ning and Fan, Heng},
	title = {Variational quantum simulation of thermal statistical states on a superconducting quantum processer},
	journal = {Chinese Physics B}
}

@article{Vijendran_2024,
	doi = {10.1088/2058-9565/ad200a},
	url = {https://doi.org/10.1088/2058-9565/ad200a},
	year = {2024},
	month = {feb},
	publisher = {IOP Publishing},
	volume = {9},
	number = {2},
	pages = {025010},
	author = {Vijendran, V and Das, Aritra and Koh, Dax Enshan and Assad, Syed M and Lam, Ping Koy},
	title = {An expressive ansatz for low-depth quantum approximate optimisation},
	journal = {Quantum Science and Technology}
}

@article{Jang:2026,
	author = {Jang, Enhyeok and Chen, Zihan and Ha, Dongho and Choi, Seungwoo and Lee, Yongju and Kwon, Jaewon and Zhang, Eddy Z. and Huang, Yipeng and Ro, Won Woo},
	date = {2026/01/26},
	doi = {10.1007/s42484-026-00357-w},
	id = {Jang2026},
	isbn = {2524-4914},
	journal = {Quantum Machine Intelligence},
	number = {1},
	pages = {5},
	title = {Layerwise retraining and freezing for multi-angle QAOA},
	url = {https://doi.org/10.1007/s42484-026-00357-w},
	volume = {8},
	year = {2026}}

@article{Araz:2024xkw,
	author = "Araz, Jack Y. and Jha, Raghav G. and Ringer, Felix and Sambasivam, Bharath",
	title = "{Thermal state preparation of the SYK model using a variational quantum algorithm}",
	eprint = "2406.15545",
	archivePrefix = "arXiv",
	primaryClass = "quant-ph",
	reportNumber = "JLAB-THY-24-4088",
	month = "6",
	year = "2024"
}

@article{Selisko_2024,
	doi = {10.1088/2058-9565/ad1340},
	url = {https://doi.org/10.1088/2058-9565/ad1340},
	year = {2023},
	month = {dec},
	publisher = {IOP Publishing},
	volume = {9},
	number = {1},
	pages = {015026},
	author = {Selisko, Johannes and Amsler, Maximilian and Hammerschmidt, Thomas and Drautz, Ralf and Eckl, Thomas},
	title = {Extending the variational quantum eigensolver to finite temperatures},
	journal = {Quantum Science and Technology}
}

@article{McClean_2016,
	doi = {10.1088/1367-2630/18/2/023023},
	url = {https://doi.org/10.1088/1367-2630/18/2/023023},
	year = {2016},
	month = {feb},
	publisher = {IOP Publishing},
	volume = {18},
	number = {2},
	pages = {023023},
	author = {McClean, Jarrod R and Romero, Jonathan and Babbush, Ryan and Aspuru-Guzik, Alán},
	title = {The theory of variational hybrid quantum-classical algorithms},
	journal = {New Journal of Physics}
}

@article{Cerezo2021,
	author = {Cerezo, M. and Arrasmith, Andrew and Babbush, Ryan and Benjamin, Simon C. and Endo, Suguru and Fujii, Keisuke and McClean, Jarrod R. and Mitarai, Kosuke and Yuan, Xiao and Cincio, Lukasz and Coles, Patrick J.},
	date = {2021/09/01},
	doi = {10.1038/s42254-021-00348-9},
	id = {Cerezo2021},
	isbn = {2522-5820},
	journal = {Nature Reviews Physics},
	number = {9},
	pages = {625--644},
	title = {Variational quantum algorithms},
	url = {https://doi.org/10.1038/s42254-021-00348-9},
	volume = {3},
	year = {2021}}

@article{PhysRevLett.125.260505,
	title = {Obstacles to Variational Quantum Optimization from Symmetry Protection},
	author = {Bravyi, Sergey and Kliesch, Alexander and Koenig, Robert and Tang, Eugene},
	journal = {Phys. Rev. Lett.},
	volume = {125},
	issue = {26},
	pages = {260505},
	numpages = {6},
	year = {2020},
	month = {Dec},
	publisher = {American Physical Society},
	doi = {10.1103/PhysRevLett.125.260505},
	url = {https://link.aps.org/doi/10.1103/PhysRevLett.125.260505}
}

@article{
	doi:10.1073/pnas.2006373117,
	author = {Guido Pagano  and Aniruddha Bapat  and Patrick Becker  and Katherine S. Collins  and Arinjoy De  and Paul W. Hess  and Harvey B. Kaplan  and Antonis Kyprianidis  and Wen Lin Tan  and Christopher Baldwin  and Lucas T. Brady  and Abhinav Deshpande  and Fangli Liu  and Stephen Jordan  and Alexey V. Gorshkov  and Christopher Monroe },
	title = {Quantum approximate optimization of the long-range Ising model with a trapped-ion quantum simulator},
	journal = {Proceedings of the National Academy of Sciences},
	volume = {117},
	number = {41},
	pages = {25396-25401},
	year = {2020},
	doi = {10.1073/pnas.2006373117},
	URL = {https://www.pnas.org/doi/abs/10.1073/pnas.2006373117}}

@article{Farhi:2016zwi,
	author = "Farhi, Edward and Harrow, Aram W.",
	title = "{Quantum Supremacy through the Quantum Approximate Optimization Algorithm}",
	eprint = "1602.07674",
	archivePrefix = "arXiv",
	primaryClass = "quant-ph",
	reportNumber = "MIT/CTP-4771, MIT/CTP-4771",
	month = "2",
	year = "2016"
}

@article{PhysRevX.10.021067,
	title = {Quantum Approximate Optimization Algorithm: Performance, Mechanism, and Implementation on Near-Term Devices},
	author = {Zhou, Leo and Wang, Sheng-Tao and Choi, Soonwon and Pichler, Hannes and Lukin, Mikhail D.},
	journal = {Phys. Rev. X},
	volume = {10},
	issue = {2},
	pages = {021067},
	numpages = {23},
	year = {2020},
	month = {Jun},
	publisher = {American Physical Society},
	doi = {10.1103/PhysRevX.10.021067},
	url = {https://link.aps.org/doi/10.1103/PhysRevX.10.021067}
}

@article{SciPostPhys.6.3.029,
	title = {Efficient variational simulation of non-trivial quantum states},
	pages = {029},
	author = {Ho, Wen Wei and Hsieh, Timothy H.},
	journal = {SciPost Phys.},
	volume = {6},
	year = {2019},
	publisher = {SciPost},
	doi = {10.21468/SciPostPhys.6.3.029},
	url = {https://scipost.org/10.21468/SciPostPhys.6.3.029}
}

@article{PhysRevA.99.052332,
	title = {Ultrafast variational simulation of nontrivial quantum states with long-range interactions},
	author = {Ho, Wen Wei and Jonay, Cheryne and Hsieh, Timothy H.},
	journal = {Phys. Rev. A},
	volume = {99},
	issue = {5},
	pages = {052332},
	numpages = {8},
	year = {2019},
	month = {May},
	publisher = {American Physical Society},
	doi = {10.1103/PhysRevA.99.052332},
	url = {https://link.aps.org/doi/10.1103/PhysRevA.99.052332}
}

@article{PRXQuantum.2.010309,
	title = {Quantifying the Efficiency of State Preparation via Quantum Variational Eigensolvers},
	author = {Matos, Gabriel and Johri, Sonika and Papi\ifmmode \acute{c}\else \'{c}\fi{}, Zlatko},
	journal = {PRX Quantum},
	volume = {2},
	issue = {1},
	pages = {010309},
	numpages = {13},
	year = {2021},
	month = {Jan},
	publisher = {American Physical Society},
	doi = {10.1103/PRXQuantum.2.010309},
	url = {https://link.aps.org/doi/10.1103/PRXQuantum.2.010309}
}

@article{Harrigan2021,
	author = {Harrigan, Matthew P. and Sung, Kevin J. and Neeley, Matthew and Satzinger, Kevin J. and Arute, Frank and Arya, Kunal and Atalaya, Juan and Bardin, Joseph C. and Barends, Rami and Boixo, Sergio and Broughton, Michael and Buckley, Bob B. and Buell, David A. and Burkett, Brian and Bushnell, Nicholas and Chen, Yu and Chen, Zijun and Ben Chiaro and Collins, Roberto and Courtney, William and Demura, Sean and Dunsworth, Andrew and Eppens, Daniel and Fowler, Austin and Foxen, Brooks and Gidney, Craig and Giustina, Marissa and Graff, Rob and Habegger, Steve and Ho, Alan and Hong, Sabrina and Huang, Trent and Ioffe, L. B. and Isakov, Sergei V. and Jeffrey, Evan and Jiang, Zhang and Jones, Cody and Kafri, Dvir and Kechedzhi, Kostyantyn and Kelly, Julian and Kim, Seon and Klimov, Paul V. and Korotkov, Alexander N. and Kostritsa, Fedor and Landhuis, David and Laptev, Pavel and Lindmark, Mike and Leib, Martin and Martin, Orion and Martinis, John M. and McClean, Jarrod R. and McEwen, Matt and Megrant, Anthony and Mi, Xiao and Mohseni, Masoud and Mruczkiewicz, Wojciech and Mutus, Josh and Naaman, Ofer and Neill, Charles and Neukart, Florian and Niu, Murphy Yuezhen and O'Brien, Thomas E. and O'Gorman, Bryan and Ostby, Eric and Petukhov, Andre and Putterman, Harald and Quintana, Chris and Roushan, Pedram and Rubin, Nicholas C. and Sank, Daniel and Skolik, Andrea and Smelyanskiy, Vadim and Strain, Doug and Streif, Michael and Szalay, Marco and Vainsencher, Amit and White, Theodore and Yao, Z. Jamie and Yeh, Ping and Zalcman, Adam and Zhou, Leo and Neven, Hartmut and Bacon, Dave and Lucero, Erik and Farhi, Edward and Babbush, Ryan},
	date = {2021/03/01},
	doi = {10.1038/s41567-020-01105-y},
	id = {Harrigan2021},
	isbn = {1745-2481},
	journal = {Nature Physics},
	number = {3},
	pages = {332--336},
	title = {Quantum approximate optimization of non-planar graph problems on a planar superconducting processor},
	url = {https://doi.org/10.1038/s41567-020-01105-y},
	volume = {17},
	year = {2021}}

@article{Du2022,
	author = {Du, Yuxuan and Huang, Tao and You, Shan and Hsieh, Min-Hsiu and Tao, Dacheng},
	date = {2022/05/23},
	doi = {10.1038/s41534-022-00570-y},
	id = {Du2022},
	isbn = {2056-6387},
	journal = {npj Quantum Information},
	number = {1},
	pages = {62},
	title = {Quantum circuit architecture search for variational quantum algorithms},
	url = {https://doi.org/10.1038/s41534-022-00570-y},
	volume = {8},
	year = {2022}}

@article{Ostaszewski2021structure,
	doi = {10.22331/q-2021-01-28-391},
	url = {https://doi.org/10.22331/q-2021-01-28-391},
	title = {Structure optimization for parameterized quantum circuits},
	author = {Ostaszewski, Mateusz and Grant, Edward and Benedetti, Marcello},
	journal = {{Quantum}},
	issn = {2521-327X},
	publisher = {{Verein zur F{\"{o}}rderung des Open Access Publizierens in den Quantenwissenschaften}},
	volume = {5},
	pages = {391},
	month = jan,
	year = {2021}
}

@article{PhysRevResearch.6.033033,
	title = {Adaptive diversity-based quantum circuit architecture search},
	author = {Huang, Yuhan and Jin, Siyuan and Zeng, Bei and Shao, Qiming},
	journal = {Phys. Rev. Res.},
	volume = {6},
	issue = {3},
	pages = {033033},
	numpages = {15},
	year = {2024},
	month = {Jul},
	publisher = {American Physical Society},
	doi = {10.1103/PhysRevResearch.6.033033},
	url = {https://link.aps.org/doi/10.1103/PhysRevResearch.6.033033}
}

@article{Alet2021,
	author = {Alet, Fabien and Hanada, Masanori and Jevicki, Antal and Peng, Cheng},
	date = {2021/02/03},
	doi = {10.1007/JHEP02(2021)034},
	id = {Alet2021},
	isbn = {1029-8479},
	journal = {Journal of High Energy Physics},
	number = {2},
	pages = {34},
	title = {Entanglement and confinement in coupled quantum systems},
	url = {https://doi.org/10.1007/JHEP02(2021)034},
	volume = {2021},
	year = {2021}}

@INPROCEEDINGS{Wang2022,
	author={Wang, Hanrui and Ding, Yongshan and Gu, Jiaqi and Lin, Yujun and Pan, David Z. and Chong, Frederic T. and Han, Song},
	booktitle={2022 IEEE International Symposium on High-Performance Computer Architecture (HPCA)}, 
	title={QuantumNAS: Noise-Adaptive Search for Robust Quantum Circuits}, 
	year={2022},
	volume={},
	number={},
	pages={692-708},
	doi={10.1109/HPCA53966.2022.00057}}

@article{Gaidai2024,
	author = {Gaidai, Igor and Herrman, Rebekah},
	date = {2024/08/14},
	doi = {10.1038/s41598-024-69643-6},
	id = {Gaidai2024},
	isbn = {2045-2322},
	journal = {Scientific Reports},
	number = {1},
	pages = {18911},
	title = {Performance analysis of multi-angle QAOA for $p > 1$},
	url = {https://doi.org/10.1038/s41598-024-69643-6},
	volume = {14},
	year = {2024}}

@inproceedings{Kulshrestha:2024qve,
	author = "Kulshrestha, Ankit and Liu, Xiaoyuan and Ushijima-Mwesigwa, Hayato and Bach, Bao and Safro, Ilya",
	title = "{QAdaPrune: Adaptive Parameter Pruning For Training Variational Quantum Circuits}",
	booktitle = "{2024 International Conference on Quantum Computing and Engineering}",
	eprint = "2408.13352",
	archivePrefix = "arXiv",
	primaryClass = "quant-ph",
	doi = "10.1109/QCE60285.2024.10264",
	month = "8",
	year = "2024"
}

@INPROCEEDINGS{Escofet2026,
	author={Escofet, Pau and Rodrigo, Santiago and Sarkar, Rohit Sarma and Almudéver, Carmen G. and Alarcón, Eduard and Abadal, Sergi},
	booktitle={2026 IEEE International Symposium on Circuits and Systems (ISCAS)}, 
	title={Quantum Circuit Pruning: Improving Fidelity via Compilation-Aware Circuit Approximation}, 
	year={2026},
	volume={},
	number={},
	pages={1521-1525},
	doi={10.1109/ISCAS66217.2026.11562331}
	}

@article{PhysRevA.110.012445,
	title = {Variational Gibbs state preparation on noisy intermediate-scale quantum devices},
	author = {Consiglio, Mirko and Settino, Jacopo and Giordano, Andrea and Mastroianni, Carlo and Plastina, Francesco and Lorenzo, Salvatore and Maniscalco, Sabrina and Goold, John and Apollaro, Tony J. G.},
	journal = {Phys. Rev. A},
	volume = {110},
	issue = {1},
	pages = {012445},
	numpages = {14},
	year = {2024},
	month = {Jul},
	publisher = {American Physical Society},
	doi = {10.1103/PhysRevA.110.012445},
	url = {https://link.aps.org/doi/10.1103/PhysRevA.110.012445}
}

@article{Robertson2026,
	author = {Robertson, Reece and Consiglio, Mirko and Stevens, Josey and Doucet, Emery and Apollaro, Tony J. G. and Deffner, Sebastian},
	date = {2026/07/11},
	doi = {10.1038/s41534-026-01318-8},
	id = {Robertson2026},
	isbn = {2056-6387},
	journal = {npj Quantum Information},
	title = {Variational Gibbs state preparation on trapped-ion devices},
	url = {https://doi.org/10.1038/s41534-026-01318-8},
	year = {2026}}

@article{He:2026utf,
	author = "He, Runhong and Liu, Chao and Hong, Xin and Chai, Qiaozhen and Zhou, Junyuan and Guan, Ji and Cui, Guolong and Ying, Shenggang",
	title = "{Hamiltonian-Aware ADAPT Variational Quantum Eigensolver for Molecular Ground-State Simulation}",
	eprint = "2606.13118",
	archivePrefix = "arXiv",
	primaryClass = "quant-ph",
	month = "6",
	year = "2026"
}

@article{PhysRevA.100.032107,
	title = {Product spectrum ansatz and the simplicity of thermal states},
	author = {Martyn, John and Swingle, Brian},
	journal = {Phys. Rev. A},
	volume = {100},
	issue = {3},
	pages = {032107},
	numpages = {10},
	year = {2019},
	month = {Sep},
	publisher = {American Physical Society},
	doi = {10.1103/PhysRevA.100.032107},
	url = {https://link.aps.org/doi/10.1103/PhysRevA.100.032107}
}
	
\end{document}